\RequirePackage{fix-cm}
\documentclass{svjour3}                     % onecolumn (standard format)
\smartqed  % flush right qed marks, e.g. at end of proof
\usepackage{color,soul}
\usepackage{multirow}
\usepackage{graphicx}
\usepackage{amsmath}

\usepackage{array}
\newcolumntype{L}{>{\centering\arraybackslash}m{3cm}}
\newcolumntype{D}{>{\centering\arraybackslash}m{2cm}}
\usepackage{mathptmx}      % use Times fonts if available on your TeX system
\usepackage{latexsym}

\usepackage{graphicx}
\usepackage{longtable}
\usepackage[ruled,vlined,linesnumbered]{algorithm2e}
\usepackage{array}
\usepackage{bookmark}

\usepackage{amsmath}
\usepackage{multirow}
\usepackage{adjustbox}
\usepackage{cases}
\usepackage{mathtools} 
\usepackage{amssymb}
\usepackage{float}
\usepackage{pgfplots,pgfplotstable}
\usepackage[bf]{subfigure}
\usepackage{setspace}
\usepackage{xcolor}
\usepackage{geometry}
\usepackage{subfigure}
\usepackage[latin1]{inputenc}
\usepackage{tikz}
\usetikzlibrary{shapes}
\usetikzlibrary{arrows}
\usepackage{multicol}
\usepackage{caption}
\usetikzlibrary{patterns}
 \journalname{Journal of Grid Computing}
\begin{document}

\title{SAF: Simulated Annealing Fair Scheduling for Hadoop Yarn Clusters}%\thanks{Grants or other notes

\author{ Mahsa Ghanavatinasab \thanks{Corresponding Author: m.ghanavatinasab@alzahra.ac.ir} \and 
	Mastaneh Bahmani \and Reza Azmi  }
%\author{Mostafa Hadadian Nejad Yousefi         \and
%        Maziar Goudarzi %etc.
%}


\institute{%M. Hadadian Nejad Yousefi \at
              Department of Computer Engineering, Alzahra University, Tehran, Iran. \\
%              \email{hadadian@ce.sharif.edu}           %  \\
}

\date{Received: date / Accepted: date}

\maketitle

\begin{abstract}
	\label{sec:1}
	Apache introduced YARN as the next generation of the Hadoop framework, providing resource management and a central platform to deliver consistent data governance tools across Hadoop clusters. Hadoop YARN supports multiple frameworks like MapReduce to process different types of data and works with different scheduling policies such as FIFO, Capacity, and Fair schedulers. DRF is the best option that uses short-term, without considering history information, convergence to fairness for multi-type resource allocation. However, DRF performance is still not satisfying due to trade-offs between fairness and performance regarding resource utilization. To address this problem, we propose Simulated Annealing Fair scheduling, SAF, a long-term fair scheme in resource allocation to have fairness and excellent performance in terms of resource utilization and MakeSpan. We introduce a new parameter as entropy, which is an approach to indicates the disorder in the fairness of allocated resources of the whole cluster. We implemented SAF as a pluggable scheduler in Hadoop Yarn Cluster and evaluated it with standard MapReduce benchmarks in Yarn Scheduler Load Simulator (SLS) and CloudSim Plus simulation framework. Finally, the results of both simulation tools are evidence to prove our claim. Compared to DRF, SAF increases resource utilization of YARN clusters significantly and decreases MakeSpan to an appropriate level.
	
	
\end{abstract}

\section{Introduction}
\label{intro}
To deal with massive volumes of data, Apache has introduced an open-source distributed data processing framework called Hadoop. In 2006 Apache introduced the first version of Hadoop. For better resource sharing and scalability, Apache later introduced the second version of Hadoop called Yet Another Resource Negotiator (YARN). Hadoop YARN improves performance in MapReduce function and scheduling over Hadoop1 via decoupling the programming model from the resource management infrastructure \cite{eadline2015hadoop}.

Yarn scheduler is in charge of distributing cluster resources among jobs in job queues. It shares resources based on the job's resource request values  \cite{vavilapalli2013apache}.Hadoop Yarn consists of three built-in schedulers: First in First Out(FIFO) Scheduler, Fair Scheduler, and Capacity Scheduler \cite{capacity}. FIFO scheduler sorts out an application in a First in First out routine in queues. With the capacity scheduler, a separately dedicated queue allows the jobs to start as soon as they are submitted. However, this is at the cost of the full cluster utilization since the capacity queue is reserved for the jobs in this queue. The Fair scheduler assigns resources to applications such that each application gets, on average, an equal dominant share of resources over time \cite{zaharia2009job}. 
There are Customized scheduling policies over default schedulers for specific applications to offer performance improvements compared to the default policies \cite{subbulakshmi2017comparison}. One of these extended policies is Dominant Resource Fairness (DRF). DRF is a fair sharing model that generalizes max-min fairness to multiple resource types \cite{ghodsi2011dominant}.
Improving the performance of a Hadoop cluster is very vital and significant for Big Data users and providers. One Contributing factor in enhancing the performance of a cluster is a superb performance scheduler. Ergo, improving the current YARN scheduler is essential for the Hadoop cluster to perform better. Another Customized scheduler is HaSTE. This scheduling technique is based on task dependency and resource demand. \cite{yao2019new} simulates the problem as a knapsack problem and uses dynamic programming to choose a set of tasks for the scheduler to start. This scheduler improves resource utilization and reduces the execution time of a given set of jobs. The shortcoming of HaSTE is that it does not guarantee fairness. Unfortunately, the fair policies implemented in YARN are memoryless, i.e., allocating resources fairly at an instant time without considering historical information. We refer to these policies as short-term policies \cite{tang2016fair}. Also, both DRF and HaSTE schedulers concentrate on short-term optimizations and apply greedy decisions. It has been mentioned that the focus on the short-term decreases their flexibility in optimizing secondary objectives \cite{grandl2016altruistic}.

Among current Hadoop policies, DRF is the only one that supports multi-resource type sharing and fairness. DRF scheduling policy is part of Hadoop schedulers and is working compatibly with the Fair share model. Fair schedulers do not offer the best performance due to fair sharing policies, so it motivates us that DRF can be further improved to have a high-performance cluster. We consider resource utilization and MakeSpan as the measurement parameters for the performance of the cluster. 
It has been proved that MapReduce task scheduling is considered as a NP-complete problem and has to be solved with proper algorithms \cite{wang2018mapreduce,yao2014lsps,moseley2011scheduling}. Simulated Annealing is a meta-heuristic optimization algorithm that has been applied to scheduling problems with multiple criteria \cite{moschakis2015multi}. Because of its simplicity in implementation and smaller computational complexity compared to other optimization schemes like genetic-algorithms and ant-colony optimization, we chose simulated annealing to proceed with our research\cite{dowsland2012simulated}.

In this paper, we present a new algorithm, Simulated Annealing Fair scheduling, SAF,  combining the entropy parameter with the DRF policy via simulated annealing. We use simulated annealing to increase the utilization of the resources in the cluster. Simulated annealing helps to avoid local maxima or minima, and DRF equalizes dominant shares of resources between users to satisfy fairness. By fairness, we merely mean that cluster resources are equitably distributed among users or applications, which is the same fairness mentioned in the DRF algorithm. We introduced a new parameter as "entropy", indicating the disorder in the fairness of allocated shares between all users. As previously mentioned, there is a trade-off between fairness and performance. To avoid this issue at the beginning of scheduling, we allocate resources to applications or users only by considering fairness. Then, the manner of resource allocation changes by leveraging fairness with the entropy parameter. When the system's fairness is at risk, DRF balances up fairness. Then, simulated annealing comes into play, which uses fitness parameter, previously presented in \cite{yao2019new} as the gap between the resource requests of tasks and the residual resource capacity of nodes, to calculate the entropy of the system to employ it to allocate resources in a high utilization manner.

Based on the results we have gained by conducting experiments using SLS \cite{sls} and CloudSim \cite{silva2017cloudsim}, SAF demonstrates an exceptional performance level and consistently exceeds the overall goals of performance like higher resource utilization of cluster, shorter MakeSpan, and allocation time cost in all experiments in comparison to default Hadoop scheduling policies.

The rest of this paper is organized as follows. We describe the background in section \ref{sec:3}. Section \ref{sec:4} presents a brief overview of customized YARN schedulers as related work. Section \ref{sec:5} introduces the framework of SAF and its components. SAF scheduling policy is explained in section \ref{sec:6}. The evaluation of the presented solution is reviewed in Section \ref{sec:7}, the conclusion is expressed in Section \ref{sec:8}, and finally for our future work, we intend to tackle our scheduling policy from different angles in section \ref{sec:9}.

\section{Background}
\label{sec:3}

With the proliferation of demands for large-scale data processing and analysis, many cluster computing frameworks were introduced. Among them, Hadoop has been attracting much attention from industry and academia. The first version of the Hadoop framework was introduced by Apache, providing a MapReduce programming model for developing programs involving big data and distributed systems. MapReduce is a computational model based on functional programming concepts of list processing. MapReduce is able to split one big job into some smaller jobs and distribute them to some nodes (slave nodes) residing in Hadoop \cite{dean2008mapreduce}. 

Apache Hadoop benefits from a Master-Slave architecture, and it comprises several components like JobTracker and TaskTracker. The JobTracker, which runs only on the master node, is responsible for cluster resource management and scheduling. TaskTracker administrates execution of map or reduce tasks on slave node \cite{ghazi2015hadoop}. JobTracker and TaskTracker are two essential processes involved in MapReduce execution in MRv1, but With the rapid growth of the demands on big data processing and analysis, Both processes are now deprecated and replaced by Resource Manager, Application Master, and Node Manager Daemons. Concisely, Apache Hadoop transforms from a monolithic to a two-layer architecture\cite{white2012hadoop}.

Hadoop offers a distributed file system entitled HDFS to store metadata and big data sets \cite{azzedin2013towards}. The HDFS works by splitting the data into file blocks and distributing it among the cluster's nodes. In fact, by using HDFS, the MapReduce framework can schedule a task where data is placed \cite{shvachko2010hadoop}. When a client submits a MapReduce application to a cluster to get processed, first, all data is distributed and replicated among processing nodes via HDFS. Second, the job is broken into multiple smaller map or reduce jobs.

\begin{figure}[h]
	\centering
	\subfigure[Hadoop 1]{
		\includegraphics[width=0.54\textwidth]{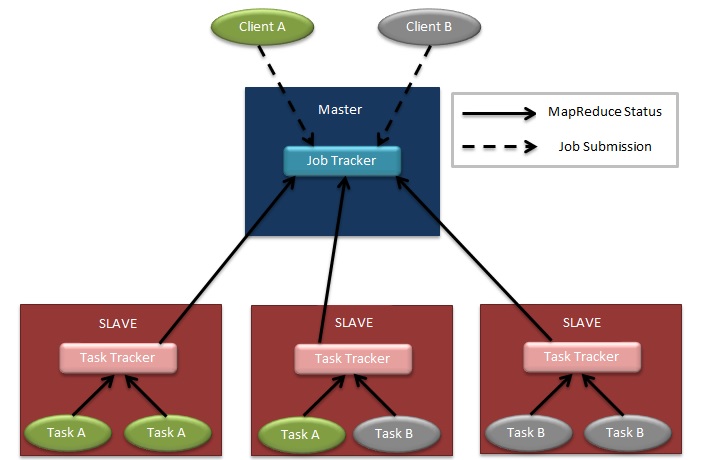}}
	\subfigure[Hadoop 2]{
		\includegraphics[width=0.54\textwidth]{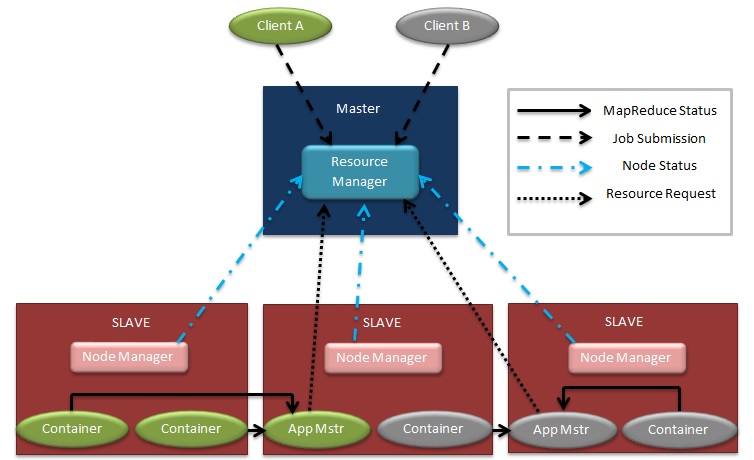}}
	\caption{Hadoop architecture evolution}
	\label{fig:architecture evolution}
\end{figure}

The evolution of Hadoop, since its creation, is tremendous. Nevertheless, the recent transition to YARN-based architecture could be considered the most significant framework upgrade since Hadoop's creation. Fig. \ref{fig:architecture evolution} depicts the evolution of Hadoop's architecture. Yet Another Resource Negotiator (YARN) expanded the Hadoop programming model from solely MapReduce to a variety of processing engines. YARN architecture for resource management provides more consistent performance, better governance of the data across Hadoop clusters, and improved security. The new resource management system is built upon the container concept. Containers, a fine-grained replacement for the slots, are resource partition abstractions with the specified size, which are used as a currency to negotiate the distribution of the Hadoop cluster's resources among the applications \cite{vavilapalli2013apache}. 

In classic Hadoop, scheduling was implemented by a centralized JobTracker, which caused performance bottleneck when cluster size increase. Also, resource management based on the slot concept lead to underutilization of the Hadoop cluster due to its coarse-grained nature \cite{vavilapalli2013apache}. Hadoop YARN separates the two main functionalities of the JobTracker in the first version of Hadoop, e.g., resource management and job scheduling/monitoring. A global ResourceManager (RM) is in charge of cluster resource management and job scheduling.Also, a per-application ApplicationMaster (AM) is required to negotiate on behalf of the jobs about requested resources with RM and monitor task execution \cite{sharma2015hadoop2}. By splitting the job coordination from RM, the YARN system is more horizontally scalable than the first generation of Hadoop. Furthermore, since RM becomes a pure scheduler, the YARN system supports different frameworks other than MapReduce, such that users can co-deploy multiple frameworks on the same cluster and choose the most suitable framework for various applications \cite{white2012hadoop}.

The performance of Hadoop clusters is sensitive to underlying scheduling. The classic Hadoop only used FIFO as scheduling policy, but with the evolution of Hadoop Yarn, other pluggable schedulers were implanted inside the ResourceManager. Users can develop and use a custom scheduler more optimized to their needs. Also, users can share Hadoop infrastructure with other users and organizations since Hadoop supports multi-tenancy in clusters \cite{capacity,fair}. Each scheduler has one or more unique scheduling policy accentuating on different Characteristics. By providing scheduling options, users are able to choose their scheduling policy well-suited to their data process and input types. The default schedulers in Hadoop Yarn are presented as follows:

\textbf{FIFO Scheduler:} In the FIFO scheduling policy, jobs are sorted out in the queue concerning the job submission order. Oldest jobs are chosen first from the queue. FIFO is simple and works well with small jobs but is not efficient with different sized jobs because larger jobs submitted first block smaller jobs to execute until they are finished \cite{vavilapalli2013apache,subbulakshmi2017comparison}. 

\textbf{Capacity Scheduler:} The Capacity scheduler shares a large multi-tenant cluster with inter-organization applications. Queues can be configured with specific users or applications and can be assigned with a fraction of cluster capacity. Queues can be provided with more resources than their real demand \cite{capacity,subbulakshmi2017comparison}.

\textbf{Fair Scheduler:} The Fair scheduling divides cluster resources between users equally unless a user has a higher priority than the others, which in that case, the user gets more resources rather than others. When a job finishes, allocated resources will be released, and then again, they are redistributed fairly amongst running applications. Fair queues can be configured with different policies such as FIFO, Fair Share, and DRF Policy \cite{fair,subbulakshmi2017comparison}.

DRF is one of the built-in policies which can be configured with the Fair scheduler. It is a Fair resource allocation policy in a system with multiple resource types demands. DRF is mostly like the fair scheduling policy for multiple resource types, except for a new parameter to determine the most demanded resource type named dominant resource. The user's dominant resource is the most demanded resource type by the instance (user or job). DRF Makes scheduling decisions by trying to equalize dominant resource usages of instances.

More precisely, DRF calculates a parameter entitled minshare ratio for each instance. Each instance has two parameters defined by the user, resource requests and, minshare. Minshare is defined as the minimum resources that are required to be allocated. If the usage of the dominant resource usage is above its minshare, then the instance is not needy. It means that the user has gotten at least its minimum shares of its demand; conversely, If the dominant resource usage is below its minshare, then the instance is needy. The DRF Selection between instance A and B in Fair queue in different situations is illustrated in Table~\ref{tab:1} as follows:
\begin{table} [!htbp]		
	\caption{DRF selection in different situation}
	\label{tab:1}  
	\centering   
	\begin{tabular}{|l|l|p{6.5cm}|}
		\noalign{\smallskip}\hline
		Instance A & Instance B & DRF Selection \\
		\hline
		Is needy & Is needy & The instance with lowest minshare ratio\\
		Is needy & Not needy & The instance that is needy \\
		Not needy & Is needy & The instance that is needy \\
		Not needy & Not needy & The instance with lowest dominant fairshare ratio \\
		\hline
	\end{tabular}
\end{table}

The minshare ratio is the result of resource usage divided by minshare (defined by the user). More precisely, according to Table~\ref{tab:variables}, $minshare\ ratio = \frac{U[DR]}{M[DR]}$. Here is an example. Assume that there are two resource types in the system, CPU and RAM, and the dominant resource for instance A is RAM. If the $M$[RAM] is 5 MegaBytes and the $U$[RAM] is 10 MegaBytes, then the $minshare\ ratio$ will be 2. Generally, the higher the minshare ratio, the higher the gap between usage and minshare. The dominant fairshare ratio is the usage divided by the cluster resource multiplied by the resource weight. Thus, $dominant\ fairshare\ ratio = \frac{U[DR]}{C[DR] . W[DR]}$. In the previous example, if the cluster's RAM is 100 MegaBytes, and the weight is 2, the dominant fairshare ratio for instance A will be 0.05. 

\begin{table} [!htbp]		
	\caption{list of DRF parameters}
	\label{tab:variables}  
	\centering   
	\begin{tabular}{|l|p{6.5cm}|}
		\noalign{\smallskip}\hline
		Variable & Description \\
		\hline
		$DR$ & dominant resource\\
		$U[R]$ & usage amount of resource R \\
		$M[R]$ & minshare amount of resource R \\
		$C[R]$ & Cluster capacity of resouce R \\
		$W[R]$ & Weight of resource R \\
		\hline
	\end{tabular}
\end{table}

\begin{figure}[h]
	\centering
	\subfigure[DRF Resource Allocation]{
		\includegraphics[width=0.42\textwidth]{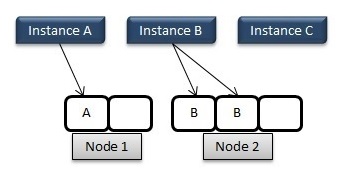}
		\label{DRF Resource Allocation}}
	\subfigure[Not DRF Resource Allocation]{
		\includegraphics[width=0.42\textwidth]{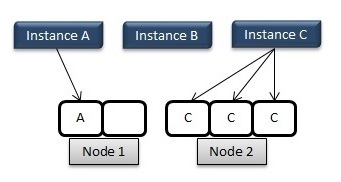}
		\label{Not DRF Resource Allocation}}
	\caption{Comparison between DRF and Not DRF Resource Allocation}
	\label{fig:Resource Allocation}
\end{figure}

We illustrate the limitation of DRF on resource utilization by a simple example. To clarify further, Consider there is only one type of resource in the cluster. Suppose, at a given moment, the system's residual capacity is 5, so that there are two unused units of resource in node 1 and 3 unused units in node 2. Also, the requirements of instance A, instance B, and instance C are 1, 2, 3 and their $dominant\ fairshare\ ratio$s are $\frac{1}{5}$,  $\frac{2}{5}$, and  $\frac{3}{5}$, respectively. Another assumption in this example is that all instances are not needy. According to the DRF policy, instance A is selected between A and B for resource allocation. After resource allocation to instance A, DRF decides whether to allocate resource to either B or C with $dominant\ fairshare\ ratio$ values $\frac{2}{5}$ and $\frac{3}{5}$, respectively. Again, DRF selects instance B with inferior value of $dominant\ fairshare\ ratio$. Thus, as shown in the Fig.~\ref{DRF Resource Allocation}, the remaining resource units in the system reduce to 2 units. However, as can be seen in the Fig.~\ref{Not DRF Resource Allocation}, if DRF policy had chosen instance C instead of instance B in the second step, the residual resource units in the system would have been equal to 1, but DRF is strict and does not have flexibility, while SAF does and gives a chance to the other option (meaning instance C here) which leads to better resource utilization. 

\section{Related Work}
\label{sec:4}
YARN provides three default schedulers. It has a pluggable scheduler policy that provides an interface to plug different implementations of schedulers. Hadoop YARN works with FIFO, Capacity, and Fair schedulers that each scheduler can apply FIFO, Fairshare, and DRF policies to their queues \cite{ghodsi2011dominant}.

There have been so many research work done on both improving performance of Hadoop using various kinds of approaches like improving data locality, reusing idle resources and modifying fair scheduling in cloud systems and data centers. Still, there has been limited work done to improve the YARN cluster's performance by improving the YARN scheduler, which is related to our work
. Therefore, we only describe researches that are relevant to improving YARN cluster performance using custom scheduling algorithms.

Since multi-resource fairness was a challenge for hierarchical scheduling in organizations, \cite{bhattacharya2013hierarchical} extends DRF for hierarchical scheduling within large organizations and defeats the old slot scheduling and solves starvation and inefficiencies, but it still needs optimization on recalculation of dominant shares in each node if the number of nodes rises up.

Tetris has been introduced as a new scheduler that uses multidimensional bin packing to achieve lower MakeSpan. It assigns the least amount of resources to the jobs with less remaining work while the jobs resource requirements can be volatile. Since fairness is not in consonance with great performance, they used heuristics to have both at the same time \cite{grandl2014multi}.

COSHH, classification, and optimization-based scheduler for heterogeneous Hadoop systems have been designed and developed at McMaster University. It aims to improve the mean completion time of jobs by reducing search overhead via optimizing search space and number of cycles and reducing communication costs, considering heterogeneity and data locality. COSHH considers the minimum requirement of jobs and satisfies fairness to prevent user starvation. To this end, it uses k-means clustering method to classify jobs. Since minshare satisfaction is more important than fairness, first, it classifies the jobs with defined minshare, and second, it repeats the classification for all jobs \cite{rasooli2014coshh}. 

In this paper \cite{pulamolu2017heuristics}, a new resource sharing algorithm called HRSYARN (Heuristics based resource sharing with fairness) is introduced, which uses Weighted Arithmetic Mean (WAM) to apply fair resource sharing between tenants. It also adds a heuristic table, which contains and updates information about resources that are lent or borrowed to and from other tenants. Their results have shown that their scheduler works better than other long-term YARN schedulers available specifically in terms of efficient resource utilization and fairness. However, there is no discussion or results in scheduling multi-type resources.

Despite most schedulers concentrating on short-term optimizations via greedy decisions Altruistic Scheduling, \cite{grandl2016altruistic} takes an approach of long-term optimization. They introduced CARBYNE, a long-term strategy that takes a fraction of the job's resources without extending their completion time. A new global resource scheduler takes leftover resources and uses them to improve cluster utilization. They showed it performs as well as DRF while reducing the completion time.

Another work that focuses on long-term fairness is \cite{tang2016fair}. To address the problem of memoryless resource allocation, they proposed Long-Term Resource Fairness (LTRF) for fair resource allocation and Hierarchical Long-Term Resource Fairness (H-LTRF) as an extension of LTRF. LTRF works only for single resource type fair allocation, which can be taken into account as its limitation.

To further improve the fair scheduling algorithm \cite{cheng2017improving} has done some modification under the restriction of fairness to the fair scheduling algorithm. They proposed a mechanism that includes several stages like job classification, pool resource allocation, job sorting, delay time adjustments, and job priority adjustments.

TMSA is a proposed schedule by \cite{wang2018mapreduce} that selects tasks in the task queue of Hadoop based on the real-time performance of the nodes and matching degree of tasks and nodes. The results are evident that TMSA reduces the completion time of batch jobs compared to existing schedulers.  

Another example of improved schedulers is HaSTE. This scheduling technique is based on Task dependency and resource demand. HaSTE scheduling is composed of two phases: initial task assignment and real-time task assignment. As soon as a resource container is released, the real-time task assignment starts. In this phase, two new parameters, urgency, fitness, and alignment, are introduced. Yao et al. states "Fitness" primarily refers to the gap between the resource demand of tasks and the residual resource capacity of nodes \cite{yao2019new}.

Furthermore, HaSTE scheduling develops an aggregation function, which is an objective function to optimize the selection of tasks based on the combination of fitness and urgency and alignment metric. HaSTE achieves higher resource utilization over time, which is because HaSTE allows jobs whose resource requests can better fit the available resource capacities to have a higher chance of getting resources and thus improves resource utilization. Also, HaSTE reduces MakeSpan of MapReduce jobs by paying heed to the correlation between the map and reduce tasks while there is a shortage of resources in the YARN cluster \cite{yao2019new}. The biggest shortcoming of HaSTE is that it does not guarantee fairness.

Based on the following experiments in evaluation (section \ref{sec:7}), our scheduling policy, SAF, shows better performance than DRF. It has proven to increase resource utilization (Memory and Vcore) by at least 90\% and reduces allocation time cost and makespan, respectively, by over 70\% and 32\%. Also, SAF outperforms HaSTE, which is evident to perform better than FIFO, Fair, and DRF, in both resource utilization (memory and core) and MakeSpan, since in the exact test environment mentioned in \cite{yao2019new}, SAF's performance vs DRF's exceeds the HaSTE's performance vs DRF's. SAF also guarantees fairness since it is based on the fair policy's main concepts.  

\section{SAF Framework}
\label{sec:5}

\subsection{Framework Structure}

Considering the importance of optimizing resource utilization, we design the SAF framework to satisfy fairness in resource allocation alongside better exploiting the cluster resources. If these two goals are met simultaneously, the consumption of resources will be more efficient. Our framework consists of two components, which we will introduce each one in detail in this section.

\subsection{Fitness Factor}
\label{sec:5.1}	
In seeking to describe the HaSTE algorithm, using the fitness \cite{yao2019new} parameter can give us a good vision of the amount of resource utilization of the cluster. As already mentioned, fitness demonstrates the gap between the resource demand of tasks and the cluster's residual resource capacity. Thus, we improve resource utilization of cluster by using fitness as a metric as:  
\begin{equation} 
F_{i}  = \sum R[i,q] . C[j,q]. W_{q}    , \forall q \in [1 , 2 , ... , k]\label{eq:fitness}
\end{equation}

The task with the highest fitness score is the best solution. $R[i,q]$ is the requested amount of resource $r_{q} $, by the $i$th task. $r_{1} , r_{2} , ... , r_{k}$ represents the k different types of resources in the system. Although there are usually only two types of resources in the system, CPU, and RAM, k types are considered for the formula to be extendable.  $C[j,q]$ is the resource capacity of type $q$ at the $j$th node, and $w_{q}$ is the weight of the $r_{q}$. Thus, the task's fitness with the number $i$ equals the multiplication of these three values. It should be noted that after the task assignment, the capacity at the node $j$ must be updated \cite{yao2019new}.

\subsection{Main Framework}

As mentioned above, DRF is an allocation policy that is used for multiple resource types. At each step, DRF picks the user with the lower dominant fairshare ratio. DRF selection is strict, but we need an algorithm that would be more flexible and, despite satisfying fairness, also improves utilization. For instance, in a situation where users are not needy, and there is little difference between users' dominant fairshare ratios, we want that the user with a higher dominant fairshare ratio also has a chance to get picked. So to attain flexibility, we decided to use a heuristic algorithm.

There are different heuristic algorithms. Many of them, which widely used as optimization techniques, are Hill-climbing, Genetic algorithm, ant-colony, and simulated annealing\cite{hillclimbingSAGA}. All four methods are easily implemented, but their performance is different for optimization problems. Generally, the Genetic algorithm has sometimes been criticized for lacking convergence proofs\cite{adaptiveSAGA}, and the Hill-climbing method might get trapped in a local minimum in the search process, which is not equal to the global one\cite{metaheuristic}. However, the Simulated Annealing algorithm provides simplicity in implementation and smaller computational complexity compared to other optimization schemes like genetic-algorithms and ant-colony optimization \cite{dowsland2012simulated}. Thus, we decided to use Simulated Annealing to test this hypothesis.

In order to satisfy fairness, we base our algorithm on the DRF policy with these modifications. Table~\ref{tab:2}~has illustrated these changes as follows:

\begin{table} [H]
	\caption{SAF selection in different situation}
	\label{tab:2}  
	\centering     
	\begin{tabular}{|l|l|p{6.5cm}|}
		\noalign{\smallskip}\hline
		Instance A & Instance B & SAF Selection \\
		\hline
		Is needy & Is needy & The instance with lower minshare ratio\\
		Is needy & Not needy & The instance that is needy \\
		Not needy & Is needy & The instance that is needy \\
		Not needy & Not needy & The instance with lower dominant fairshare ratio with a probability to choose the other one \\
		\hline
	\end{tabular}
\end{table}

As can be seen from Table~\ref{tab:2}, the conditions in our solution are the same as the conditions in DRF policy, only a few changes have been made. As mentioned above, the dominant fairshare ratio is calculated by the usage divided by the cluster resources multiplied by the resource weight. According to the simulated annealing, the probability metric should be calculated and used to give a chance to the instances that are less needy to dodge local minima in users' fairshares. So, we define probability as:
\begin{equation} 
p = e ^  \frac{-\triangle fairshare}{T}
\label{eq:probabality}
\end{equation}
The numerator of the fraction expresses the difference between dominant fairshare ratios of instance A and instance B. The denominator, annealing temperature, investigates the total fitness entropy of the system to improve resource utilization. In the next part, we first describe the methods of temperature calculation and annealing strategies and then present the overview of the SAF algorithm in the next section.
\label{here:1}
\subsection{Temperature Estimator}
\label{sec:Temperature Estimator}
Since we decided to use simulated annealing to optimize resource utilization, we should introduce the methods of calculating the entropy as temperature estimators. 

Entropy is a parameter to measure the existent disorder in a system \cite{nourani1998comparison}. In our policy, the disorder depends on the fitness parameters, as previously defined. So, we need a one-parameter entropy for a discrete random variable. There are various estimators to get entropy. We provide a rushed overview of eclectic one-parameter entropies below.

\subsubsection{Shannon entropy}

One of the most popular entropies is Shannon's rate-distortion function. Shannon defined $H(P) = - K  \sum  p_{i} . \log  p_{i} $ as the entropy of $P$ (a set of $p_{i}$, the Probability Density Function of data values) where $K > 0$ \cite{shannon1948mathematical}.	
\subsubsection{Renyi entropy}
\begin{figure} [!htbp]
	\centering
	\includegraphics[height=0.48\textwidth,width=0.75\textwidth]{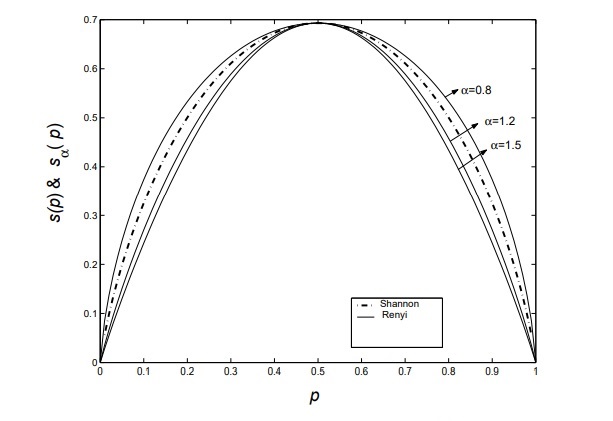}
	\caption{Shannon and Renyi Entropy Functions \cite{dukkipati2006generalized}}
	\label{fig:shannon-vs-Renyi}
\end{figure}

The Renyi entropy is a spectrum of generalizations to the Shannon and other Entropies defined by $H_{\alpha} \left(P \right)$ $ =  \frac{K}{1 - \alpha} \log (\sum p_{i} ^ \alpha )$. $H_{\alpha}  \left(P \right)$ as a function of $\alpha$, can be regarded as a measure of the entropy of $P$ where $P$ equals to the finite discrete probability distribution of $p_{i}$ in $P$ where $\alpha > 0$, $ \alpha \neq 1$ \cite{renyi1961measures}.  
Fig.~\ref{fig:shannon-vs-Renyi} shows the comparison of Shannon entropy function and Renyi entropy function for various values of $\alpha$ \cite{dukkipati2006generalized}.

Specific cases of Renyi entropy are as follows:
Hartley entropy ($\alpha = 0$), Shannon entropy ($\alpha = 1$), Collision entropy ($\alpha = 2$), and min entropy ($\alpha = \infty$) \cite{crooks2017measures}. If $K \geq 1$, Renyi entropy turns in to other forms of entropies as follows:

\begin{itemize}
	
	\item\textbf{Hartley entropy} 
	is obtained by $H_{0}  \left(P \right)$ $ = \log n$, the logarithm of the cardinality which is the number of the elements of $P$ \cite{crooks2017measures}. In fact, when $\alpha$ approaches zero, the entropy is computed regardless of the probabilities.
	
	\item\textbf{Collision entropy}
	is the Renyi entropy of order 2 which is denoted through $H_{2}  \left(P \right)$ $ = -\log(\sum p_{i} ^ 2 )$ formula \cite{crooks2017measures}. The Renyi's quadratic entropy is less than the Shannon entropy except for the uniform distributions where the two are equal\cite{bennett1992experimental}.
	\item\textbf{Min entropy}
	is Renyi entropy when $\alpha$ approaches infinity and is calculated by $H_{\infty}  \left(P \right)$ $= - \log (\max p_{i})$ \cite{crooks2017measures}, which means that this estimator is increasingly determined by the events with the highest probability. In fact, it is called min-entropy since it's the smallest entropy in the Renyi family.
\end{itemize} 		
\subsubsection{Tsallis entropy}
\begin{figure} [!htbp]
	\centering
	\includegraphics[height=0.5\textwidth,width=0.75\textwidth]{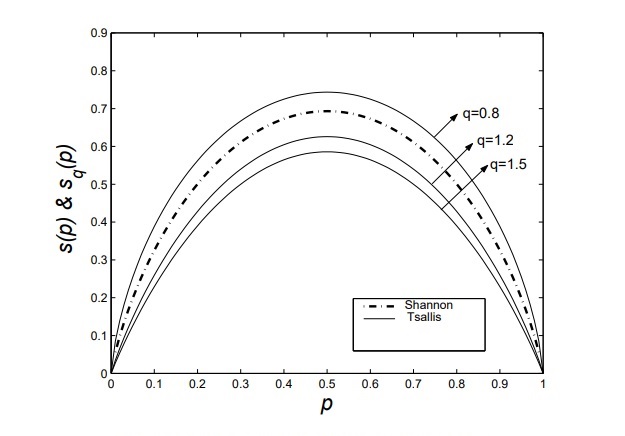}
	\caption{Shannon and Tsallis Entropy Functions \cite{dukkipati2006generalized}}
	\label{fig:shannon-vs-tsallis}
\end{figure}

is another kind of one-parameter entropy which is gained by $H_{q}$ $ = \frac{K}{q - 1} (1 - \sum p_{i} ^ q)$ \cite{crooks2017measures}, where $q > 0$. q is called the nonextensive index. As $q$ approaches 1, Tsallis estimator approaches Shannon estimator. Fig.~\ref{fig:shannon-vs-tsallis} shows the comparison of Shannon entropy function and Tsallis entropy function for various values of $q$ \cite{dukkipati2006generalized}.

It should be noted that in all estimators, $K$ is a positive constant, which usually equals Boltzmann's constant, and each of them is the entropy of the set of probabilities $p_{1}$,...,$p_{n}$.

It is evident that Renyi entropy more focuses on significant events with greater $\alpha$; conversely, events with smaller probability are Less affected. In other words, the larger the $\alpha$, the higher the effect of the bigger $p_{i}$ than the smaller ones. This is true for Tsallis entropy too. In that way, this entropy will increase the effect of the larger probability by using higher $q$. Meanwhile, we need an entropy that the influence of each $p_{i}$ on its behavior is the same as others. So, our choice is to estimate entropy by Shannon's. Hence We define T as the temperature of the system as:
\begin{equation} 
T = - k  \sum  p_{i} . \log  p_{i}  , \forall i \in [1,\infty]
\label{eq:temperature}
\end{equation}

Where $p_{i}$ is the probability distribution function of the fitness score in Eq.\ref{eq:fitness} of running tasks in the cluster.

To make sure that Shannon entropy is appropriately selected, we will examine all
mentioned entropies in the SAF algorithm and express the result in the evaluation section.  

\subsection{Annealing Strategy}

The cooling method is one of the most critical parts of the simulated annealing algorithm to get executed more effectively. Because when the temperature of the system rises, the cooling can balance up the whole system. If the temperature cools down too fast, the algorithm converges to a local minimum. Conversely, if it is reduced too slow, simulated annealing will be convergent slowly and won't be trapped in local minima \cite{du2016simulated}. Some of the approaches of the Geometric cooling schedule are presented below. 

\begin{itemize} 
	\item\textbf{Exponential schedule}
	If the initial temperature is $T_{0}$, then for each temperature cycle $k$, the temperature is cooled down with the following formula:
	$T_{k} = T_{0} . \alpha ^ k$ where $0.8 \leq \alpha \leq 0.9$ \cite{kirkpatrick1983optimization,nourani1998comparison}.
	
	\item\textbf{linear schedule}
	This method decreases the temperature by dividing the initial temperature $T_{0}$ by one plus the result of the multiplication of $\alpha$ factor and the number of the temperature cycle. So, $T_{k} = \frac{T_{0}}{1 + \alpha . k}$ where $ \alpha > 0$ \cite{kirkpatrick1983optimization,nourani1998comparison}.
	
	\item\textbf{logarithmic schedule}
	is introduced by Geman and Geman\cite{geman1984stochastic}. In this scheme, the initial temperature $T_{0}$ is divided by, one plus the result of multiplication of $\alpha$ and logarithm of temperature cycle $K$ plus one. More precisely, $T_{k} = \frac{T_{0}}{1 + \alpha . \log (1 + k)}$ where $ \alpha > 0$ \cite{nourani1998comparison,mascagni1989review}.
	
	\item\textbf{Quadratic schedule}
	The declined temperature is computed by the initial temperature $T_{0}$ divided by one plus the logarithm of one plus the temperature cycle $k$ squared. Thus, the progression converges to $T_{k} = \frac{T_{0}}{1 + \alpha . \log (1 + k ^ 2)}$ where $ \alpha > 0$ \cite{martin2009comparison} and .       
\end{itemize} 

\begin{figure} [!htbp]
	\centering
	\includegraphics[height=0.45\textwidth,width=0.75\textwidth]{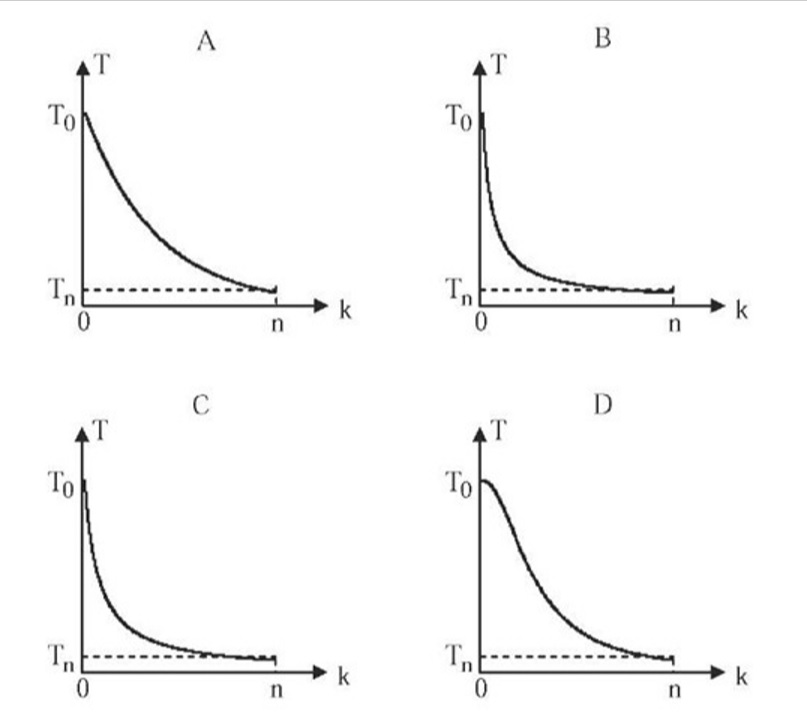}
	\caption{Cooling Curves: (A) Exponential, (B) logarithmic, (C) linear, (D) quadratic \cite{martin2009comparison}}
	\label{fig:cooling curves}
\end{figure}

According to  Fig.~\ref{fig:cooling curves}, the standard exponential cooling strategy can escape from local minima because of a mild slope in the graph's initial and central parts and softer at the final part of the graph. Moreover, the standard error of the iteration's count is better in quadratic and exponential methods because the tails of these two curves cancel out faster than others \cite{martin2009comparison}. Hence, our choice for cooling strategy is the exponential manner. However, we test all cooling strategies with the SAF scheduler and report the result in the evaluation section\ref{cloudsimplusResult}.

\section{SAF Scheduler}
\label{sec:6}

To address the problems presented before, we propose SAF, Simulated Annealing Fair scheduling, as a novel allocation policy for multiple resource types in the Hadoop ecosystem. HaSTE \cite{yao2019new} has tended to focus on improving resource utilization of the cluster rather than the fairness of resource allocation, which is playing a vital factor in Hadoop scheduling policy performance. Fitness \cite{yao2019new} and the dominant fairshare ratio of users \cite{ghodsi2011dominant}, two determinant factors in job scheduling between two users, have been employed in SAF to satisfy fairness in resource allocation, and improve resource utilization. At each step for every user, DRF calculates the share of each resource type allocated to that user. The maximum share amongst different types of resource requests of a user is called that users' dominant share. The resource type affiliated with the dominant share is called the dominant resource. Also, if the dominant share is divided by the weight parameter, the result is the dominant fairshare ratio. DRF allocates the resources to the user with the lowest dominant fairshare ratio.	

SAF uses Eq.~\ref{eq:fitness} to calculate the fitness score of a user, selected to be scheduled for resource allocation. As mentioned in the previous section, SAF consists of two crucial components, DRF and simulated annealing Based on users' shares of resources. SAF executes either simulated annealing fair scheduling or DRF, depending on whether users' shares of resources are greater than their minshares or not (being Needy). By using DRF, we are confident that each users' shares of resources will be fair. So, the first and foremost factor in SAF scheduling is fairness. It is worth noting that SAF, just like DRF and Fairshair model \cite{fair}, compares two instances, $i$ and $i + 1$, according to its policy.

Algorithm~\ref{alg:saf} and Fig.~\ref{fig:flowchart} show that in the situation in which both instances are not needy, SAF checks up on fairness through the difference of two instances' dominant fairshare ratios. SAF employs simulated annealing to maintain the fairness through $\triangle fairshare$ and improves resource utilization of cluster by using the entropy of job's fitnesses as a second factor to calculate the temperature for simulated annealing, where $\triangle fairshare$ denotes the gap between $fairshare (i+1)$ and $fairshare (i)$ as follows:

\begin{equation} 
\triangle fairshare = fairshare (i+1) - fairshare (i)
\end{equation}
\\
\\
\begin{multicols}{2}
	{\begin{algorithm}[H]
			\DontPrintSemicolon
			\SetAlgoLined		
			\BlankLine		
			Check $i$ is needy or not\\
			Check  $i+1$ is needy or not \\
			\BlankLine
			\If{$i$ is needy \textup{\bf \&\&}  $i+1$ is needy}{		
				pick instance with lower minshare ratio	}
			\ElseIf{$i$ is not needy \textup{\bf \&\&} $i+1$ is needy}{		
				pick instance $i+1$ }		    
			\ElseIf{$i$ is needy \textup{\bf \&\&} $i+1$ is not needy}{
				pick instance $i$ }
			\ElseIf{ $i$ is not needy \textup{\bf \&\&} $i+1$ is not needy}
			{
				$T = - k  \sum  p_{i} . (\log  p_{i} )  , \forall i \in [1,\infty]$ \\	 	
				\While{cluster is not full}
				{
					\If{$\triangle fairshare < 0$}
					{
						pick instance $i+1$
					}
					\Else{$rand=Random(0,1)$
						
						\uIf{$ p = e ^ \frac{-\triangle fairshare}{T}  > rand $} 
						{pick instance $i+1$
						}
						\uElse{pick instance $i$}					
						\uIf{$T > cold\_temperature$}
						{
							$T = cooling\_rate . T $
						}
						\uElse{
							$T = - k  \sum  P_{i} . (\log  P_{i} )  , \forall i \in [1,\infty]$
						}
						
					}
					
				}				            
			}    
			
			\caption{SAF}
			\label{alg:saf}
		\end{algorithm}
	}
	{% Define block styles
		\tikzstyle{decision} = [diamond, draw, aspect=2, fill=blue!10, 
		text width=9em, text badly centered, node distance=3cm, inner sep=0pt]
		\tikzstyle{value} = [trapezium, trapezium left angle=70, trapezium right angle=-70, draw, fill=olive!20, text width=6em, text badly centered, node distance=3cm]
		\tikzstyle{block} = [rectangle, draw, fill=red!10, 
		text width=10em, text centered, rounded corners, minimum height=2em]
		\tikzstyle{line} = [draw, -latex']
		\tikzstyle{cloud} = [draw, ellipse, text width=4em, text badly centered, fill=green!10, minimum height=2em] 
		\scalebox{0.5}{    
			\begin{tikzpicture}[auto]
			% Place nodes
			\node [cloud] (start) {start};
			\node [value, below of=start, node distance=2cm] (input) {Input Value of Two Instances $i$ and $i+1$};
			\node [decision, below of=input] (needy1) {IF $i = needy$ AND $i+1 = needy$};   
			\node [decision, below of=needy1] (needy2) {IF $i \ne needy$ AND $i+1 = needy$};
			\node [decision, below of=needy2] (needy3) {IF $i = needy$ AND $i+1 \ne needy$};    
			\node [block, below of=needy3, node distance=2cm] (firstT) { $T = - k  \sum  p_{i} . (\log  p_{i} )$ };
			\node [decision, below of=firstT, node distance=2cm] (cluster) {IF Cluster is Full};    
			\node [decision, below of=cluster] (fairshare) {IF $\triangle fairshare < 0$};
			\node [block, below of=fairshare, node distance=2cm] (rand) { $rand=Random(0,1)$ };
			\node [decision, below of=rand, node distance=2cm] (random) {IF $ p = e ^ \frac{-\triangle fairshare}{T}  > rand $};
			\node [value, below of=random] (picki) {Pick Instances $i$};
			\node [decision, below of=picki] (coldT) {IF $T > cold\_temperature$};
			\node [block, below of=coldT, node distance=2cm] (secondT) { $T = - k  \sum  p_{i} . (\log  p_{i} )$ };
			\node [value, right of=needy1, node distance=5cm] (lowerminshareratio) {Pick Instance with Lower minshare ratio};
			\node [value, right of=needy2, node distance=5cm] (picki2) {Pick Instance $i$};
			\node [value, right of=needy3, node distance=5cm] (picki3) {Pick Instance $i+1$};
			\node [value, right of=fairshare, node distance=6cm] (picki4) {Pick Instance $i+1$};
			\node [value, right of=random, node distance=6cm] (picki5) {Pick Instance $i$};    
			\node [block, right of=coldT, node distance=6cm] (multiT) { $T = cooling\_rate . T $ };
			\node [cloud, right of=picki3, node distance=4cm] (stop) {stop};    
			% Draw edges
			%, node distance=2cm
			\path [line] (start) -- (input);  
			\path [line] (input) -- (needy1);     
			\path [line] (needy1)-- node [near start] {Yes}(lowerminshareratio.west);
			\path [line] (lowerminshareratio.east) -| (stop);
			\path [line] (needy1) -- node {No}(needy2);
			\path [line] (needy2) -- node {No}(needy3);
			\path [line] (needy2)-- node [near start] {Yes}(picki2.west);
			\path [line] (picki2.east) -| (stop);
			\path [line] (needy3) -- node {No}(firstT);
			\path [line] (needy3)-- node [near start] {Yes}(picki3.west);
			\path [line] (picki3) -- (stop);
			\path [line] (firstT) -- (cluster);
			\path [line] (cluster.east)-| node [near start]{Yes}(stop);
			\path [line] (cluster) -- (fairshare);    
			\path [line] (fairshare)-- node [near start] {Yes}(picki4.west);
			\path [line] (fairshare) -- node {No}(rand);  
			\path [line] (rand) -- (random);
			\path [line] (random) -- node {No}(picki);
			\path [line] (random)-- node [near start] {Yes}(picki5.west); 
			\path [line] (picki) -- (coldT);
			\path [line] (coldT) -- node {No}(secondT);
			\path [line] (coldT)-- node [near start] {Yes}(multiT.west);
			% \path [line] (picki4.east) -| (cluster.west);   
			\path [line] (picki4.east) -| ([yshift=-1cm,xshift=+7cm] secondT.east) -- ([yshift=-1cm, xshift=-2cm] secondT.west)coordinate (aux) |- (cluster.west);
			\path [line] (picki5.east) -| ([yshift=-1cm,xshift=7cm] secondT.east) -- ([yshift=-1cm, xshift=-2cm] secondT.west)coordinate (aux) |- (cluster.west);
			\path [line] (multiT.east) -| ([yshift=-1cm,xshift=+7cm] secondT.east) -- ([yshift=-1cm, xshift=-2cm] secondT.west)coordinate (aux) |- (cluster.west);
			\path [line] (secondT.east) -| ([yshift=-1cm,xshift=+7cm] secondT.east) -- ([yshift=-1cm, xshift=-2cm] secondT.west)coordinate (aux) |- node [near end] {No}(cluster.west);
			% \path [line] (picki) --(stop);  
			
			\end{tikzpicture}}
		
		\captionof{figure}{SAF Flowchart}
		\label{fig:flowchart}	
	}
\end{multicols}

\bigskip	
If $\triangle fairshare$ is negative, then the amount of $fairshare (i+1)$ is less than $fairshare (i)$. So, SAF picks out instance$(i+1)$. But if $\triangle fairshare$ is positive, then the value of $fairshare (i+1)$ is greater than the value of $fairshare (i)$, thus SAF can picks instance$(i+1)$ with the probability $p$.		 

As shown in Eq.~\ref{eq:probabality}, the probability $p$ depends on $\triangle fairshare$ and temperature $T$. The smaller the change in fairness ($\triangle fairshare$) and the higher the temperature( entropy of fitness in the cluster), the more likely it is for the algorithm to accept the next solution (instance $i+1$). 

The temperature is calculated in the first cycle of simulated annealing by Eq.~\ref{eq:temperature}. The constant cooling \_rate cools down each cycle's temperature until it reaches the cold\_temperature. As soon as the temperature reaches cooling\_rate, it is recalculated by Eq.~\ref{eq:temperature}. This time, recent fitnesses will be applied in calculating the temperature.

The mechanism of SAF scheduling is fully shown in the following example. In this example, the cluster's total capacity is $\langle100\,CPU, 100\,GB\rangle$, which means that the cluster offers 100 CPUs and 100 GigaByte of RAM to its users. User A and user B have requested $\langle30\,CPU, 10 \,GB\rangle$ and $\langle20\,CPU, 40 \,GB\rangle$ of resources, respectively. The dominant resource of user A is CPU, and the dominant resource of user B is RAM. The minshare parameter defined by users, is $\langle15\,CPU, 2\,GB\rangle$ for user A and $\langle10\,CPU,  20\,GB\rangle$ for user B. In this example, DRF allocates resources to user A and B according to Table~\ref{tab:example}.

\begin{table} [H]
	\caption{DRF Scheduling Example }
	\centering
	\renewcommand{\arraystretch}{1.5}
	\resizebox{0.9\textwidth}{!}{
		\begin{tabular}{|c|c|c|c|c|c|c|c|}

			\hline
			\multirow{2}{*}{\textbf{Iteration}} & \multirow{2}{*}{\textbf{Schedule}}  & \multicolumn{2}{c|}{\textbf{User A}} & \multicolumn{2}{c|}{\textbf{User B }}& \multirow{2}{*}{\textbf{ CPU total.alloc}} & \multirow{2}{*}{\textbf{ RAM total.alloc}}\\ 
			% \hline
			% \textbf{Inactive Modes} & \textbf{Description}\\
			\cline{3-6}
			& & \textbf{res.shares} & \textbf{dom.shares} & \textbf{res.shares} & \textbf{dom.shares} & & \\
			%\hhline{~--}
			\hline
			1 & B & $\langle 0 , 0 \rangle$ & 0 &  $\langle \frac{2}{10} , \frac{4}{10} \rangle$ &
			$\frac{4}{10}$ & $\frac{2}{10}$ & $\frac{4}{10}$ \\ \hline
			
			2 & A & $\langle \frac{3}{10} , \frac{1}{10} \rangle$ & $\frac{3}{10}$ &  $\langle \frac{2}{10} , \frac{4}{10} \rangle$ &	$\frac{4}{10}$ & $\frac{5}{10}$ & $\frac{5}{10}$ \\ \hline \hline
			
			3 & A & $\langle \frac{6}{10} , \frac{2}{10} \rangle$ & $\frac{6}{10}$ &  $\langle \frac{2}{10} , \frac{4}{10} \rangle$ &	$\frac{4}{10}$ & $\frac{8}{10}$ & $\frac{6}{10}$ \\ \hline 
			
			4 & B & $\langle \frac{6}{10} , \frac{2}{10} \rangle$ & $\frac{6}{10}$ &  $\langle \frac{4}{10} , \frac{8}{10} \rangle$ &	$\frac{8}{10}$ & $\frac{10}{10}$ & $\frac{10}{10}$ \\ \hline 
			
		\end{tabular}
	}			
	\label{tab:example}
\end{table}

If this example was performed by SAF, until the third iteration of scheduling, SAF selection would have been the same as DRF selection since SAF would have chosen DRF as scheduling policy. However, In the third iteration, for both users, $resource\,usage\,ratios > minshare\,ratios$, so they are not needy anymore; hence SAF would have chosen simulated Annealing over DRF.  Simulated annealing would have calculated $\triangle fairshare\,of\,user\,B\,and\,A >0 $, and then it would have picked out user B with a probability P. Below is the proof that SAF definitely would have picked user B in the third cycle of scheduling.

\begin{equation*}
F_{i}  = \sum R[i,q] . C[j,q]. W_{q}    , \forall q \in [CPU,RAM] 
\end{equation*}		
$  W_{CPU} = W_{RAM} = 1$ :  
\begin{equation*} 
\begin{rcases*}
R[B,CPU] = 20 \\
R[B,RAM] = 40 \\
C[cluster,CPU] = 50 \\
C[cluster,RAM] = 50 \\
\end{rcases*} 	F_{B} = 20 \times 50 + 40 \times 50 =3000 
\end{equation*}
and similary,
\begin{equation*}
\begin{rcases*}
R[A,CPU] = 30 \\
R[A,RAM] = 10 \\
C[cluster,CPU] = 50 \\
C[cluster,RAM] = 50 \\
\end{rcases*} 	F_{A} = 30 \times 50 + 10 \times 50 =2000 
\end{equation*}
Since at this point only two tasks have been scheduled: $n=2 ,\: p_{A} =  p_{B} = 0.5 $ 

\begin{equation*}
\begin{array}{l}
T_{0} =  -k \sum p_{i}. \log p_{i} \Longrightarrow 
\\T_{0} = -K(p_{A}.\log p_{A} + p_{B}.\log p_{B}) \xrightarrow{K=1000} 
T_{0} = -1000 \times \log (0.5)= 301 
\\
P = e^ \frac{-\triangle fairshare}{T0} = e^ \frac{0.4 - 0.3}{301} \simeq 1 \;\blacksquare
\end{array}
\end{equation*}

%%%%%%%%%%%%%%%%%%%%%%%%%%%%%%%%%%%%%%%%%%%%%%%%%%%%%%%%%%%%%%%%%%%%%%%%
\iffalse
\[
\left.
\begin{drcases}
\text{for } 0 \leq n \leq 1 \\
\text{for } 0 \leq n \leq 1 \\
\text{for } 0 \leq n \leq 1  \\
\end{cases}
\right\} = xy
\]
\[
X(m,n) = \left\{\begin{array}{lr}
x(n), & \text{for } 0\leq n\leq 1\\
x(n-1), & \text{for } 0\leq n\leq 1\\
x(n-1), & \text{for } 0\leq n\leq 1
\end{array}\right\} = xy
\]

Suppose that in the middle of the process, user A and user B occupy $\langle15/3\:CPU, 5/1\:GB\rangle$ and $\langle10.2\:CPU, 2.4\:GB\rangle$ of total resources, respectively. Both users are not needy since the usage ratio of their dominant resource are above the minshare ratio of their dominant resource. In this situation, DRF calculates the fairshare ratios for each user. According to Table 3, DRF picks user A. However, the SAF has another scheduling. SAF calculates fairshare of dominant resource and fitness for each user and gives a chance to the user with greater fairshare by probability p. So, SAF selects user B. Thus, in addition to having  fairness in our policy, we have better performance in resource utilization.	
\fi
%%%%%%%%%%%%%%%%%%%%%%%%%%%%%%%%%%%%%%%%%%%%%%%%%%%%%%%%%%%%%%%%%%%%%%%%

\section{Evaluation}
\label{sec:7}
We evaluate the performance of SAF by conducting experiments using two different tools that can simulate large-scale clusters. These two simulators employed in these experiments are SLS\cite{sls} and CloudSim Plus \cite{silva2017cloudsim}. In the following subsections, the characteristics of each experiment are presented.

\subsection{Use of SLS Simulator}
\label{slsResult}

The Yarn Scheduler Load Simulator designed by Apache is a tool to prototype a new scheduler and simulate its behavior and performance without a real Hadoop cluster \cite{sls}. It emulates YARN components and all the communication heartbeat messages between them within a single Java Virtual Machine (JVM) \cite{elshater2015study}. We implemented SAF as an extension to DRF policy and compared their performance to each other. The performance metrics used in this experiment are considered as MakeSpan of a batch of MapReduce jobs and resource utilization of the Hadoop yarn cluster.

The characteristics of these experiments and configurations of YARN are exactly the same as the HaSTE scheduler, so our experiment's results would be totally and logically comparable to HaSTE. We tested SAF with 8, 16, 32, 64 and 128 nodes. Each node has a capacity of 8GB memory and 8 virtual CPU cores, i.e., $\langle 8\:vcores , 8\:GB\rangle$. We added SAF policy to Hadoop resource manager v3.1.0 and replaced it on the master node. Then, We used a workload that consists of four Wordcount jobs. A 3.5GB text file was used as an input to be parsed, and the HDFS block size was set to 64MB. The configuration of each job and the resource requests are shown in Table~\ref{tab:sls}.

\begin{table} [H]
	\caption{Workload Configuration}
	\label{tab:sls} 
	\label{tab:5}  
	\centering     
	\begin{tabular}{|>{\centering\arraybackslash}p{1.1cm}|>{\centering\arraybackslash}p{0.8cm}|>{\centering\arraybackslash}p{1.1cm}|>{\centering\arraybackslash}p{2.5cm}|>{\centering\arraybackslash}p{2.8cm}|}
		\noalign{\smallskip}\hline
		Job ID & Map & Reduce & Resource requests of Map & Resource requests of Reduce \\ 
		\hline
		1 & 31 & 5 &  $\langle 2\:vcores , 1\:GB \rangle$ & $\langle 2\:vcores 1\:GB,  \rangle$ \\ \hline
		2 & 31 & 5 &  $\langle 3\:vcores , 1\:GB \rangle$ & $\langle 2\:vcores, 1\:GB  \rangle$ \\
		\hline
		3 & 31 & 5 &  $\langle 4\:vcores , 1\:GB \rangle$ & $\langle 3\:vcores, 1\:GB  \rangle$ \\
		\hline
		4 & 31 & 5 &  $\langle 5\:vcores , 1\:GB \rangle$ & $\langle 3\:vcores, 1\:GB  \rangle$ \\
		\hline
		\noalign{\smallskip}
	\end{tabular}
\end{table}

\begin{figure}[H]  %figure 7
	
	\subfigure[\bf Core usage over time (core per second)]{
		\begin{tikzpicture}
		%here
		\pgfplotsset{height=0.37\textwidth,width=0.51\textwidth,compat=newest}
		\begin{axis}[
		xtick=data,ytick={0,10,20,30,40,50,60,70},symbolic x coords={8,16,32,64,128},
		label style={/pgf/number format/1000 sep=},
		ylabel={Core usage over time (Coreps)},
		xlabel= {Number of nodes}, 
		legend style={at={(0.5,1.2)},anchor=north,legend columns=-1,font=\scriptsize},
		enlarge x limits=0.15,
		ybar,ymin=0,bar width=7pt,]
		\addplot [pattern=horizontal lines, pattern color=blue] 
		coordinates {(8,20.92) (16,21.43) 	(32,21.19) 
			(64,28.23) (128,28.15)  };
		\addplot  [pattern=north east lines, pattern color=purple]
		coordinates {(8,57.25) (16,43.67) 	(32,40.32) 
			(64,53.38) (128,62.35)  };
		\legend{DRF,SAF}
		\end{axis}
		\end{tikzpicture}
		
	}
	%%%%%%%%%%%%%%%%%%%%%%%%%%%%%%%%%%%%%%%%%%%%%%%%%%%%%%%%%%%%%%%5
	\subfigure[\bf Memory usage over time (GigaByte per second)]{ 
		
		\begin{tikzpicture}
		%here
		\pgfplotsset{height=0.37\textwidth,width=0.51\textwidth,compat=newest}
		\begin{axis}[
		xtick=data,ytick={0,10,20,30,40,50,60,70},symbolic x coords={8,16,32,64,128},
		label style={/pgf/number format/1000 sep=},
		ylabel={Memory usage over time (GBps)},
		xlabel= {Number of nodes}, 
		legend style={at={(0.5,1.2)},anchor=north,legend columns=-1,font=\scriptsize},
		enlarge x limits=0.15,
		ybar,ymin=0,bar width=7pt,]
		\addplot [pattern=horizontal lines, pattern color=blue] 
		coordinates {(8,22.36) (16,21.17) (32,21.23) 
			(64,28.56) (128,28.86) };
		\addplot [pattern=north east lines, pattern color=purple]
		coordinates {(8,56.9) (16,42.16491228) 	(32,40.92) 
			(64,52.66) (128,63.40) };
		\legend{DRF,SAF}
		\end{axis}
		\end{tikzpicture}
	}
	%%%%%%%%%%%%%%%%%%%%%%%%%%%%%%%%%%%%%%%%%%%%%

	\caption{Average resource utilization of SAF and DRF under the workload of four Wordcount jobs.}
	\label{fig:SLSusage}
\end{figure}
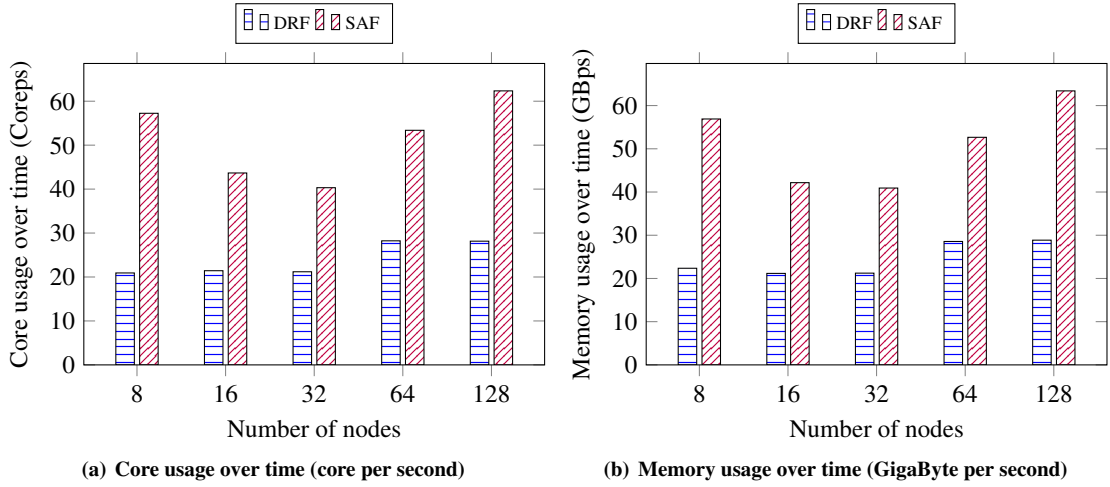

Fig.~\ref{fig:SLSusage} shows our tests result with different numbers of nodes in the SLS environment. The Memory-Usage and Vcore-Usage over time in SAF scheduler are much higher than DRF resource usage. It is trivial that both of these values have increased by more than 90\% in all cases, based on the percentage increase formula calculated by $\frac{amount\, of \, change}{initial \, amount} \times100 $. So at best state, in a 8-node cluster, memory-usage over time and vcore-usage over time have increased by 154\% and 174\%, respectively. Also with a 64-node cluster, we have the smallest increment in these two parameters. More precisely, in this case, memory-usage and vcore-usage over time have raisen by 90.14\% and 93.68\%.

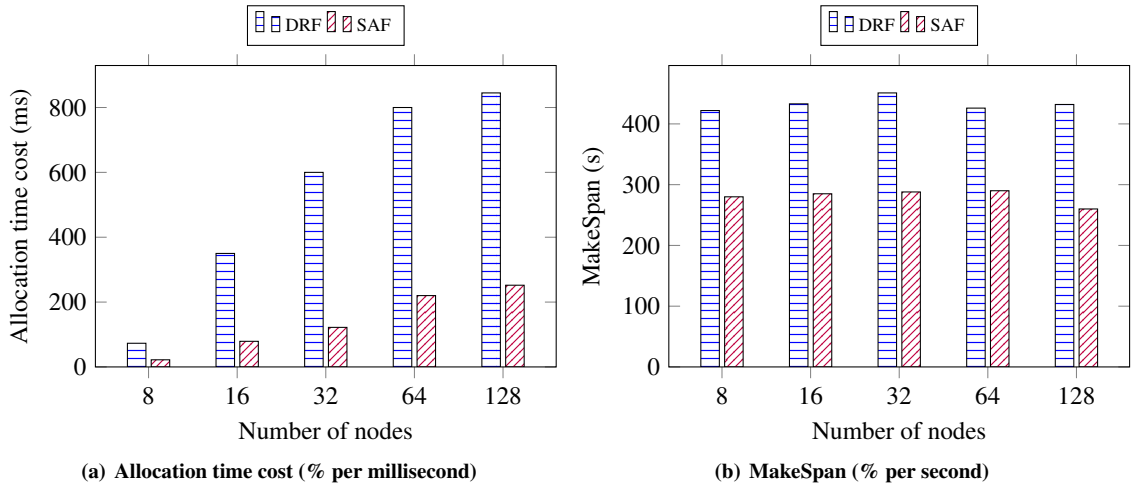
\begin{figure}[H]  		%figure 8

	\subfigure[\bf Allocation time cost (\% per millisecond)]{
		%here
		\begin{tikzpicture}
		\pgfplotsset{height=0.37\textwidth,width=0.51\textwidth,compat=newest}
		\begin{axis}[
		xtick=data,ytick={0,200,400,600,800,1000},symbolic x coords={8,16,32,64,128},
		label style={/pgf/number format/1000 sep=},
		ylabel={Allocation time cost (ms)},
		xlabel= {Number of nodes}, 
		legend style={at={(0.5,1.2)},anchor=north,legend columns=-1,font=\scriptsize},
		enlarge x limits=0.15,
		ybar,ymin=0,bar width=7pt,]

		\addplot [pattern=horizontal lines, pattern color=blue] 
		coordinates {(8,73) (16,350) (32,600) 
			(64,800) (128,845)};
		\addplot [pattern=north east lines, pattern color=purple]
		coordinates {(8,22) (16,79)(32,122) 
			(64,220) (128,252) };
		\legend{DRF,SAF}
		\end{axis}
		\end{tikzpicture}
	}
	%--------------------------------------------------------------------%
	\subfigure[\bf MakeSpan (\% per second)]{
		%here
		\begin{tikzpicture}
		\pgfplotsset{height=0.37\textwidth,width=0.51\textwidth,compat=newest}
		\begin{axis}[
		xtick=data,ytick={0,100,200,300,400,500},symbolic x coords={8,16,32,64,128},
		label style={/pgf/number format/1000 sep=},
		ylabel={MakeSpan (s)},
		xlabel= {Number of nodes}, 
		legend style={at={(0.5,1.2)},anchor=north,legend columns=-1,font=\scriptsize},
		enlarge x limits=0.15,
		ybar,ymin=0,bar width=7pt,]
		
		\addplot [pattern=horizontal lines, pattern color=blue] 
		coordinates {(8,422) (16,433) (32,451) 
			(64,426) (128,432)  };
		\addplot  [pattern=north east lines, pattern color=purple]
		coordinates {(8,280) (16,285) (32,288) 
			(64,290) (128,260) };
		\legend{DRF,SAF}
		\end{axis}
		\end{tikzpicture}
		
	}
	
	\caption{Allocation time cost and MakeSpan of SAF and DRF under the workload of four wordcount jobs.}	
	\label{fig:SLSTime}	
\end{figure}

In addition to the resource utilization, we investigate the metrics related to time like Allocation Time Cost and MakeSpan. The required time to allocate the resources to jobs is allocation time cost, and the MakeSpan is the distance in time that elapses from the start of work to the end. These two performance metrics are compared in Fig.~\ref{fig:SLSTime}. Not only MakeSpan but also allocation time cost has been significantly reduced in the SAF scheduler. The MakeSpan of jobs has dropped by over 32\%, and the allocation time cost has declined by over 70\%. 

Therefore the SLS results show that SAF increases resource utilization by at least 90\% and decreases allocation time cost and MakeSpan of MapReduce jobs in yarn cluster by more than 70\% and 32\%, respectively in comparison to DRF policy. The characteristics of this experiment are specified as the same as HaSTE scheduler under the simple workload experiment, and since HaSTE only improves resource utilization of cluster up to 40\% over DRF, we are assured that SAF outperforms HaSTE scheduler in the manner of resource utilization of YARN cluster \cite{yao2019new}.   		 

\subsection{Use of CloudSim Plus Simulator}
\label{cloudsimplusResult}

To compare DRF and SAF scheduling policies, we also conducted an experiment using CloudSim Plus, developed to model and simulate Cloud Computing Infrastructures and Services. CloudSim Plus, which is an extension of CloudSim framework \cite{calheiros2011cloudsim}, simulates virtualized Datacenters environments consists of VMs, memory, storage, and bandwidth. It provides provisioning of hosts to VMs, Allocating tasks to VMs, and monitoring application execution \cite{belalem2010approaches}. In the following, We state our specific experimental configurations.

There is one data center with one host to avoid CPU overhead for VM migration and performance degradation. Since the processor cores' total processing capacity (in Million Instructions per Second) is equally distributed among the VMs that are residing in hosts, the minimum amount of CPU capacity of the host equals the multiplication of the number of VMs and the CPU capacity of each VM. The same is true for the amount of memory, storage, and bandwidth. It is important to note that each VM's CPU capacity can be different, but here, we consider them identical. 

We used PlanetLab data as the input of the cloudlets, in which the length of each of them equals 10000. Each task has its own resource request and minshare. Each resource request consists of the number of CPUs and the amount of RAM that the corresponding task requires. The CPU and RAM requests for each task are in the range [1,8] and [100, 2000].  

The VMs are configured with 1000 MIPS, 2000 MB memory, 10 GB storage, 1 GB/s bandwidth, and 16 PE (Processing Element). So, as mentioned earlier, the host is minimally configured with (16 $\times$ number of VMs) CPU, (1000 $\times$ number of CPUs) MIPS CPU speed, (2000 $\times$ number of VMs) MB RAM, (10 $\times$ number of VMs) GB storage and (1 $\times$ number of VMs) GB/s bandwidth.

In our experiment, we set task number n = {50, 150, 250, 350, 450, 650, 850, 1050}. The test results in this environment are shown in the following Figures. Fig.~\ref{fig:usageovertime} depicts that CPU and RAM usages have significantly increased. 

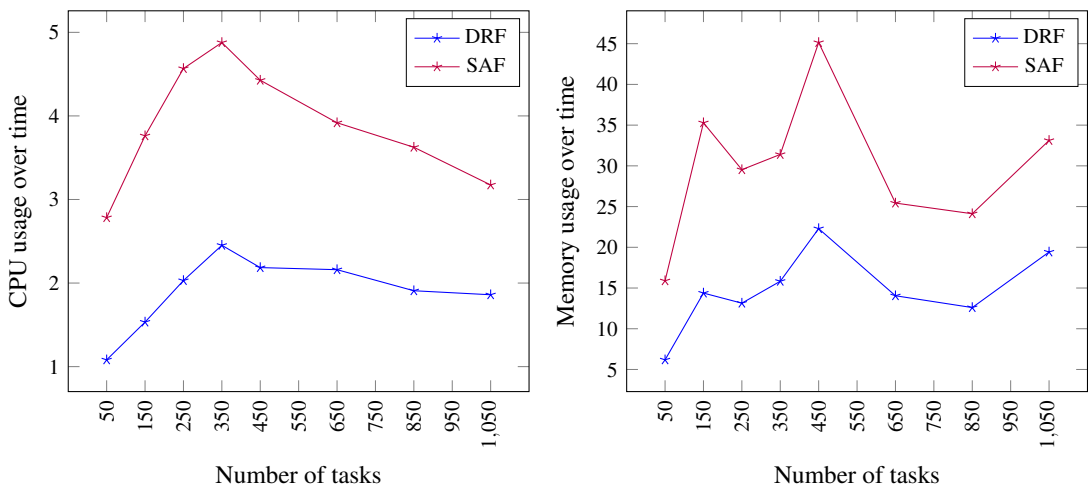
\begin{figure}[H]		%figure 7
	\centering
	
	\subfigure[\bf CPU usage of SAF VS. DRF over time (\% per millisecond)]{
		\begin{tikzpicture}
		\pgfplotsset{width=0.51\textwidth,compat=newest,every tick label/.append style={font=\small}}
		\begin{axis}[
		xtick={50,150,...,1050}, ytick={0,1,...,6},xticklabel style={rotate=90},
		xlabel={Number of tasks},  
		ylabel={CPU usage over time},
		legend style={line width=0.2pt,font=\small}
		]
		\addplot [blue,mark=star] 
		coordinates {(50,1.081) (150,1.533) (250,2.03) (350,2.452) (450,2.185) (650,2.16)  (850,1.908) (1050,1.86)};
		
		\addplot [purple,mark=star]
		coordinates {(50,2.783) (150,3.762) (250,4.567) (350,4.878) (450,4.427) (650,3.918)  (850,3.625) (1050,3.174)};
		\legend{DRF,SAF}
		\end{axis}
		\end{tikzpicture}
	}
	%--------------------------------------------------------------------%
	\subfigure[\bf Memory usage of SAF VS. DRF over time (\% per millisecond)]{
		\begin{tikzpicture}
		\pgfplotsset{width=0.51\textwidth,compat=newest,every tick label/.append style={font=\small}}
		\begin{axis}[
		xtick={50,150,...,1050}, ytick={0,5,...,50},xticklabel style={rotate=90},
		xlabel={Number of tasks},  
		ylabel={Memory usage over time},
		legend style={line width=0.2pt,font=\small}
		]
		\addplot [blue,mark=star] 
		coordinates {(50,6.18) (150,14.38) (250,13.14) (350,15.83) (450,22.29) (650,14.05)  (850,12.6) (1050,19.42)};
		
		\addplot [purple,mark=star]
		coordinates {(50, 15.89) (150,35.29) (250,29.53) (350,31.4) (450,45.14) (650,25.43)  (850,24.12) (1050,33.12)};
		
		\legend{DRF,SAF}
		\end{axis}
		\end{tikzpicture}
	}
	\caption{Ressource usages over time in SAF VS. DRF Schedulers}
	\label{fig:usageovertime}
	
\end{figure}

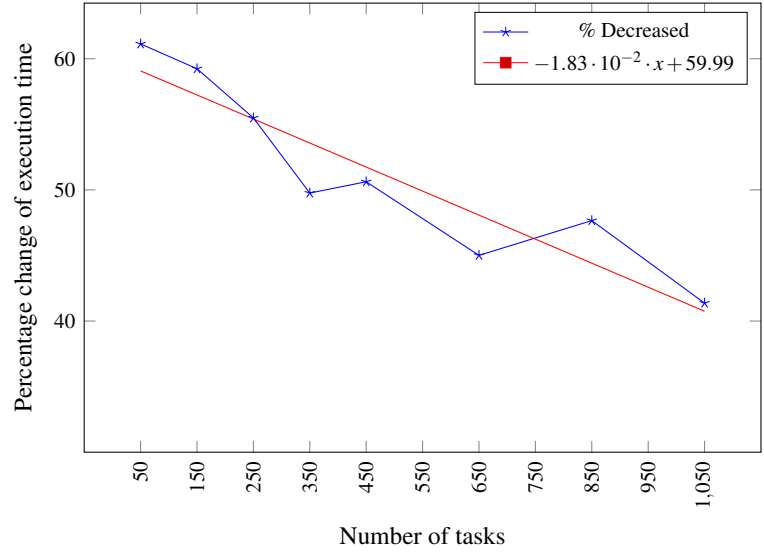
\begin{figure} [H]  % figure 8
	\centering
	\begin{tikzpicture} 
	\pgfplotsset{height=0.5\textwidth,width=0.7\textwidth,compat=newest,every tick label/.append style={font=\small}}
	\begin{axis}[
	xtick={50,150,...,1050},xticklabel style={rotate=90}, ytick={40,50,...,80} ,
	xlabel={Number of tasks},  
	ylabel={Percentage change of execution time},
	legend style={line width=0.2pt,font=\small},
	ymin=30
	]
	\addplot [blue,mark=star] 
	coordinates {(50,61.13) (150,59.24) (250,55.49) (350,49.76) (450,50.62) (650,45.01) (850,47.66) (1050,41.37)};
	\addplot table[
	y={create col/linear regression={y=Y}},mark=none] % compute a linear regression from the input table
	{
		X 		Y
		50 		61.13 
		150		59.24
		250		55.49
		350		49.76
		450		50.62
		650		45.01
		850		47.66
		1050	41.37
	};
	%\xdef\slope{\pgfplotstableregressiona} %<-- might be handy occasionally
	
	\legend{ \% Decreased, $\pgfmathprintnumber{\pgfplotstableregressiona} \cdot x  
		\pgfmathprintnumber[print sign]{\pgfplotstableregressionb}$}
	\end{axis}
	\end{tikzpicture}
	\caption{Decreased Percentage of Execution Time SAF VS. DRF}
	\label{fig:cloudtime}
\end{figure}

At best case with 50 tasks, both CPU and RAM usages over time have increased by 157\% in SAF scheduler, which means that these values in SAF are more than 2.57 times higher than DRF values. In the worst case with 1050 tasks, both CPU usage over time and RAM usage over time have increased by more than 70\%, which means they are 1.7 times higher in SAF scheduler.
The execution time is equal to the interval between the starting time of scheduling until the end of the simulation when all cloudlets are finished. Finally, the execution time values are given in Fig.~\ref{fig:cloudtime}. 
As is seen, the SAF scheduling policy results have decreased relative to DRF in all cases. The reduction is above 41\%. The least execution time reduction has occurred in the case with 1050 tasks, which has decreased from 2395 ms in DRF to 1404 ms in SAF. Also, we see the most reduction in execution time occurs in the case with 50 tasks, which is 61.13\%. The red line shows the regression line of the indicated data.

\begin{figure} [H]		%figure 9
	
	\subfigure[\bf CPU usage of Shannon VS. other entropies over time (\% per millisecond)]{   		%cpu over time
		\begin{tikzpicture} 
		\pgfplotsset{height=0.3\textwidth,width=0.51\textwidth,compat=newest,every tick label/.append style={font=\small}}
		\begin{axis}[
		xtick={50,150,...,1050},xticklabel style={rotate=90}, ytick={1,2,3,4,5,6},
		xlabel={Number of tasks},  
		ylabel={CPU usage over time},
		legend style={line width=0.2pt,font=\small}
		]
		\addplot [blue,mark=star] 
		coordinates {(50,2.78) (150,3.76) (250,4.56) (350,4.67) (450,4.42) (650,3.91) (850,3.65) (1050,3.17)};
		
		\addplot [red,mark=star]
		coordinates {(50,3.12) (150,3.82) (250,4.82) (350,4.98) (450,4.39) (650,4.03) (850,4.08) (1050,3.2)};

		\addplot [green,mark=star]
		coordinates {(50,2.45) (150,3.5) (250,4.38) (350,4.54) (450,4.16) (650,3.81) (850,3.75) (1050,2.96)};
		
		\addplot [purple,mark=star]
		coordinates {(50,3) (150,3.8) (250,4.7) (350,4.77) (450,4.41) (650,4.09) (850,3.95) (1050,3.24)};
		
		\addplot [pink,mark=star]
		coordinates {(50,2.78) (150,3.53) (250,4.58) (350,4.52) (450,4.25) (650,4.18) (850,3.75) (1050,2.94)};
		
		\addplot [orange,mark=star]
		coordinates {(50,2.75) (150,3.46) (250,4.62) (350,4.54) (450,4.27) (650,4.09) (850,3.75) (1050,2.98)};
		
		\legend{}
		\end{axis}
		\end{tikzpicture}
		\label{fig:cpu-entropy}
	}
	%--------------------------------------------------------------------%
	\subfigure[\bf Memory usage of SAF using Shannon VS. other entropies over time (\% per millisecond)]{
		\begin{tikzpicture} %memory over time
		\pgfplotsset{height=0.3\textwidth,width=0.51\textwidth, compat=newest,every tick label/.append style={font=\small}}
		\begin{axis}[
		xtick={50,150,...,1050}, ytick={0,5,...,50},xticklabel style={rotate=90},
		xlabel={Number of tasks},  
		ylabel={Memory usage over time},
		legend style={at={(0.5,1.2)},anchor=north,legend columns=-1,font=\small}
		]
		\addplot [blue,mark=star] 
		coordinates {(50,15.89) (150,35.29) (250,29.53) (350,30.08) (450,45.14) (650,25.43) (850,24.12) (1050,33.12)};
		
		\addplot [red,mark=star]
		coordinates {(50,17.85) (150,35.91) (250,31.2) (350,32.11) (450,44.83) (650,26.21) (850,26.95) (1050,33.43)};
		
		\addplot [green,mark=star]
		coordinates {(50,14.01) (150,32.81) (250,28.36) (350,29.28) (450,42.49) (650,24.78) (850,24.77) (1050,30.94)};
		
		\addplot [purple,mark=star]
		coordinates {(50,17.15) (150,35.7) (250,30.4) (350,30.73) (450,45.04) (650,26.55) (850,26.1) (1050,33.87)};
		
		\addplot [pink,mark=star]
		coordinates {(50,15.89) (150,33.17) (250,29.65) (350,29.11) (450,43.34) (650,27.13) (850,24.79) (1050,30.73)};
		
		\addplot [orange,mark=star]
		coordinates {(50,15.7) (150,32.46) (250,29.9) (350,29.28) (450,43.63) (650,26.59) (850,24.8) (1050,31.12)};
		
		\legend{}
		\end{axis}
		\end{tikzpicture}
		\label{fig:mem-entropy}
	}
	%--------------------------------------------------------------------%
	\subfigure[\bf Execution time of SAF using Shannon VS. other entropies over time (ms)]{ %execution time

		\begin{tikzpicture} 
		
		\pgfplotsset{height=0.3\textwidth,width=0.51\textwidth,compat=newest,every tick label/.append style={font=\small}}
		\begin{axis}[
		xtick={50,150,...,1050}, ytick={0,200,...,1600},xticklabel style={rotate=90},
		xlabel={Number of tasks},  
		ylabel={Execution time(ms)},
		legend style={at={(1.75,0.75)},anchor=north,legend columns=1},
		legend style={line width=0.2pt,font=\small}
		]
		\addplot [blue,mark=star] 
		coordinates {(50,82) (150,172) (250,243) (350,332) (450,433) (650,733) (850,1029) (1050,1404)};
		
		\addplot [red,mark=star]
		coordinates {(50,73) (150,169) (250,230) (350,311) (450,436) (650,711) (850,921) (1050,1391)};
		
		\addplot [green,mark=star]
		coordinates {(50,93) (150,185) (250,253) (350,341) (450,460) (650,752) (850,1002) (1050,1503)};
		
		\addplot [purple,mark=star]
		coordinates {(50,76) (150,170) (250,236) (350,325) (450,434) (650,702) (850,951) (1050,1373)};
		
		\addplot [pink,mark=star]
		coordinates {(50,82) (150,183) (250,242) (350,343) (450,451) (650,687) (850,1001) (1050,1513)};
		
		\addplot [orange,mark=star]
		coordinates {(50,83) (150,187) (250,240) (350,341) (450,448) (650,701) (850,1001) (1050,1494)};
		
		\legend{Shannon,Hartley,Collision,Min,Tsallis (q=0.8),Tsallis (q=1.2)}
		\end{axis}
		\end{tikzpicture}
		\label{fig:exetime-entropy}
		
	}
	\caption{Resource usage over time and execution time of SAF using different entropies}
	\label{fig:entropies}
\end{figure}
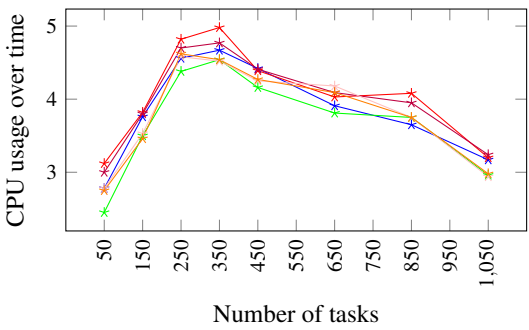
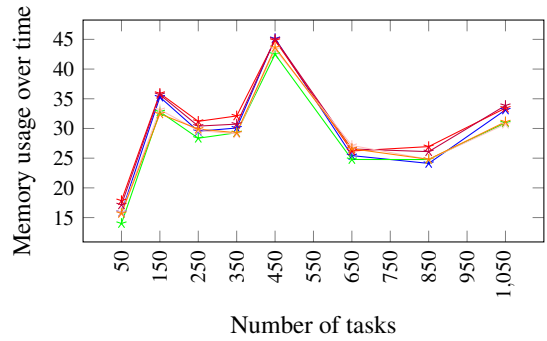
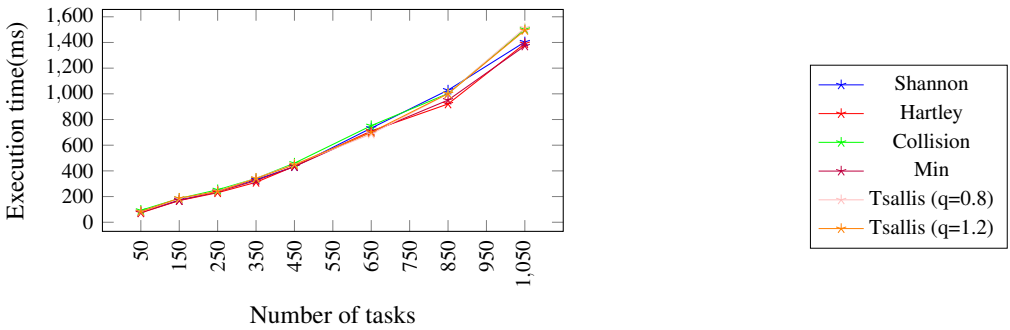

To make sure that we chose the correct entropy as a temperature estimator, we tested the rest of the entropies and compared them with Shannon. The results revealed that all of them were unobtrusive on the final results. As shown in Fig.~\ref{fig:cpu-entropy} and Fig.~\ref{fig:mem-entropy}, the results almost overlap, since they had a modest effect on the CPU and RAM usages over time. Also, the execution time graphs of SAF using various entropies almost overlap in Fig.~\ref{fig:exetime-entropy}, and only slight differences can be observed at runtime. Thus choosing the basic entropy, Shannon, is correct.

\begin{figure} [H]  %figure 10

	\subfigure[\bf CPU usage of SAF using Exponential VS. other annealing strategies (\% per millisecond)]{   	% cpu over time 
		\begin{tikzpicture} 
		\pgfplotsset{height=0.3\textwidth,width=0.51\textwidth,compat=newest,every tick label/.append style={font=\small}}
		\begin{axis}[
		xtick={50,150,...,1050},xticklabel style={rotate=90}, ytick={1,2,3,4,5,6},
		xlabel={Number of tasks},  
		ylabel={CPU usage over time},
		legend style={line width=0.2pt,font=\small}
		]
		\addplot [blue,mark=star] 
		coordinates {(50,2.78) (150,3.76) (250,4.56) (350,4.67) (450,4.42) (650,3.91)  (850,3.65) (1050,3.17)};
		
		\addplot [purple,mark=star]
		coordinates {(50,2.81) (150,2.88) (250,4.06) (350,3.87) (450,4.39) (650,3.50)  (850,3.29) (1050,2.73)};
		
		\addplot [green,mark=star]
		coordinates {(50,2.68) (150,3.31) (250,4.36) (350,4.53) (450,4.34) (650,3.87)  (850,3.52) (1050,2.9)};
		
		\addplot [orange,mark=star]
		coordinates {(50,2.48) (150,2.91) (250,3.89) (350,3.65) (450,3.88) (650,3.34)  (850,3.1) (1050,2.62)};
		
		\legend{}
		\end{axis}
		\end{tikzpicture}
		\label{fig:cpu-annealing}
	}
	%--------------------------------------------------------------------%
	\subfigure[\bf Memory usage of SAF using Exponential VS. other annealing strategies (\% per millisecond)]{
		\begin{tikzpicture} %memory over time
		\pgfplotsset{height=0.3\textwidth,width=0.51\textwidth, compat=newest,every tick label/.append style={font=\small}}
		\begin{axis}[
		xtick={50,150,...,1050}, ytick={0,5,...,50},xticklabel style={rotate=90},
		xlabel={Number of tasks},  
		ylabel={Memory usage over time},
		legend style={at={(0.5,1.2)},anchor=north,legend columns=-1,font=\small}
		]
		\addplot [blue,mark=star] 
		coordinates {(50,15.89) (150,35.29) (250,29.53) (350,30.08) (450,45.14) (650,25.43)  (850,24.12) (1050,33.12)};
		
		\addplot [purple,mark=star]
		coordinates {(50,16.09) (150,27.09) (250,26.28) (350,24.97) (450,44.83) (650,22.76)  (850,21.75) (1050,28.53)};
		
		\addplot [green,mark=star]
		coordinates {(50,15.33) (150,31.12) (250,28.25) (350,29.20) (450,44.32) (650,25.12)  (850,23.30) (1050,30.31)};
		
		\addplot [orange,mark=star]
		coordinates {(50,14.16) (150,27.34) (250,25.18) (350,23.55) (450,39.65) (650,21.72)  (850,20.53) (1050,27.42)};
		
		\legend{}
		\end{axis}
		\end{tikzpicture}
		\label{fig:mem-annealing}
	}
	%--------------------------------------------------------------------%
	\subfigure[\bf Execution time of SAF using Exponential VS. other annealing strategies (ms)]{ %execution time

		\begin{tikzpicture} 
		
		\pgfplotsset{height=0.3\textwidth,width=0.51\textwidth,compat=newest,every tick label/.append style={font=\small}}
		\begin{axis}[
		xtick={50,150,...,1050}, ytick={0,200,...,1600},xticklabel style={rotate=90},
		xlabel={Number of tasks}, 
		ylabel={Execution Time(ms)},
		legend style={at={(1.8,0.5)},anchor=north,legend columns=1},
		legend style={line width=0.2pt,font=\small}
		]
		\addplot [blue,mark=star] 
		coordinates {(50,82) (150,172) (250,243) (350,332) (450,433) (650,733) (850,1029) (1050,1404)};
		
		\addplot [purple,mark=star]
		coordinates {(50,81) (150,224) (250,273) (350,400) (450,436) (650,819)  (850,1141) (1050,1630)};
		
		\addplot [green,mark=star]
		coordinates {(50,85) (150,195) (250,254) (350,342) (450,441) (650,742)  (850,1065) (1050,1534)};
		
		\addplot [orange,mark=star]
		coordinates {(50,92) (150,222) (250,285) (350,424) (450,493) (650,858)  (850,1209) (1050,1696)};
		
		\legend{Exponential,Linear,Logarithmic,Quadratic}
		\end{axis}
		\end{tikzpicture}
		\label{fig:exetime-annealing}
	}
	\caption{Usages over time and execution time of various annealing strategies}
	\label{fig:coolingstrategies}
\end{figure}
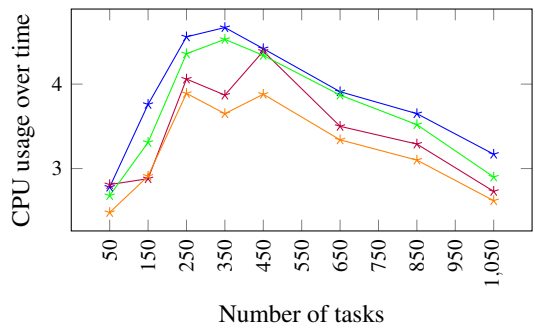
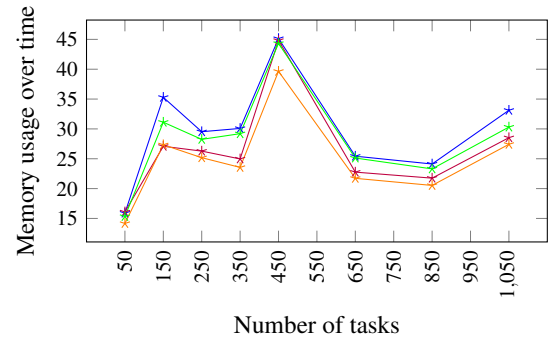
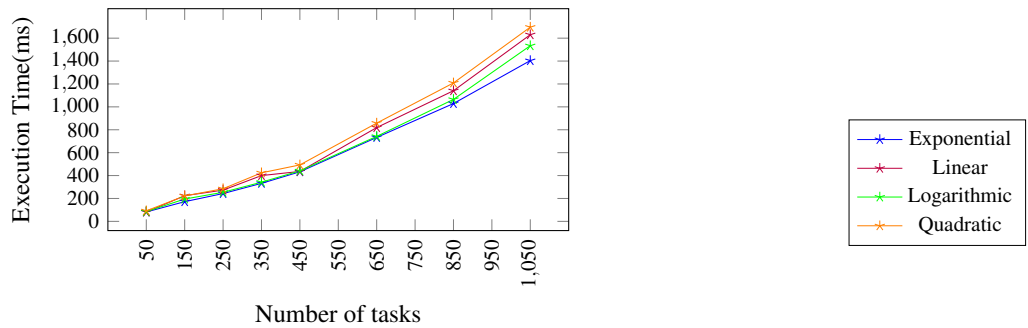

Furthermore, to ensure that the annealing strategy is adequately selected, we decided to test all of the strategies described in section ~\ref{sec:Temperature Estimator}. The outcome shows that choosing any strategy, including Exponential, Linear, Logarithmic, and Quadratic, does not make much difference in the experiment's final result. For example, as shown in Fig.~\ref{fig:cpu-annealing} and Fig.~\ref{fig:mem-annealing}, the usages over time graphs are very close. Also, in Fig.~\ref{fig:exetime-annealing}, the execution time graphs in Exponential, Logarithmic, and Quadratic are almost identical, only the runtime of Exponential strategy has decreased slightly. So, the Exponential strategy is a suitable choice.

Our test results in the CloudSim Plus environment depict that SAF policy reduces execution time by at least 41\%. Also, the resource utilization increases significantly. This value in the SAF is at least 1.7 times higher than the DRF resource utilization, which means that SAF has a higher performance in resource utilization.

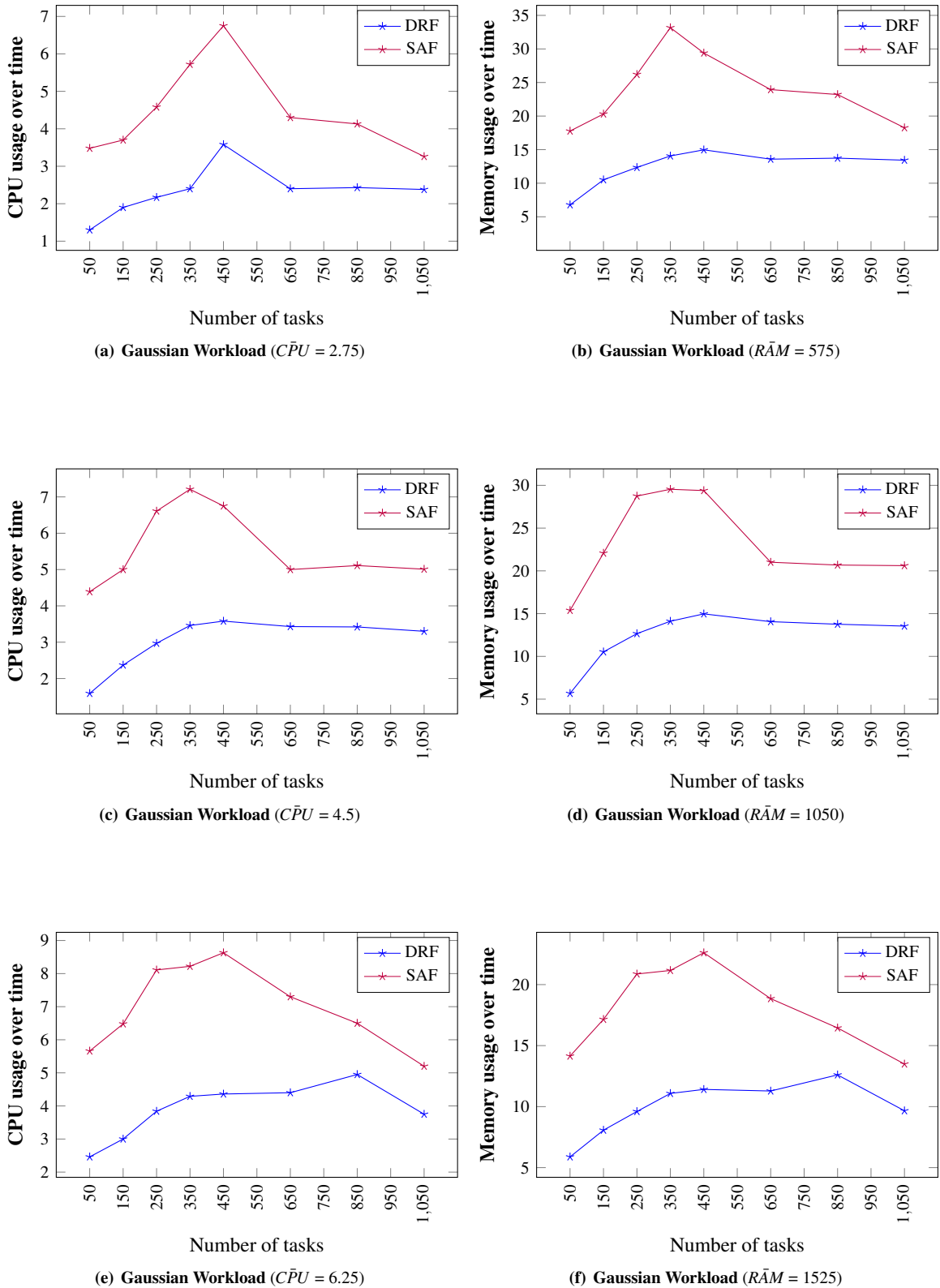
\begin{figure} [H] 	%figure 11
	\centering
	\subfigure[\bf Gaussian Workload \normalfont ($\bar{CPU}$ = 6.25)]{
		\begin{tikzpicture}  % C-1 cpu
		\pgfplotsset{height=0.35\textwidth,width=0.51\textwidth,compat=newest,every tick label/.append style={font=\small}}
		\begin{axis}[
		xtick={50,150,...,1050}, ytick={1,2,...,9},xticklabel
		style={rotate=90}, 
		xlabel={Number of tasks},  ylabel=\bf{CPU usage over time},
		legend style={line width=0.2pt,font=\small}
		]
		
		\addplot [blue,mark=star] 
		coordinates {(50,2.46) (150,3) (250,3.84) (350,4.29) (450,4.36) (650,4.4)  (850,4.95) (1050,3.75)};
		
		\addplot [purple,mark=star]
		coordinates {(50, 5.66) (150,6.48) (250,8.11) (350,8.22) (450,8.63) (650,7.3)  (850,6.5) (1050,5.2)};
		\legend{DRF,SAF}
		\end{axis}
		\end{tikzpicture}
		\label{fig:gaussian3-cpu}
	}
	%--------------------------------------------------------------
	\subfigure[\bf Gaussian Workload \normalfont ($\bar{RAM}$ = 1525)]{
		\begin{tikzpicture} C-2 memory
		\pgfplotsset{height=0.35\textwidth,width=0.51\textwidth,compat=newest,every tick label/.append style={font=\small}}
		\begin{axis}[
		xtick={50,150,...,1050}, ytick={5,10,...,25},xticklabel style={rotate=90},
		xlabel={Number of tasks},  ylabel=\bf{Memory usage over time},
		legend style={line width=0.2pt,font=\small}
		]
		
		\addplot [blue,mark=star] 
		coordinates {(50,5.88) (150,8.07) (250,9.6) (350,11.08) (450,11.41) (650,11.28)  (850,12.6) (1050,9.66)};
		
		\addplot [purple,mark=star]
		coordinates {(50,14.14) (150,17.15) (250,20.87) (350,21.16) (450,22.61) (650,18.85)  (850,16.45) (1050,13.49)};
		
		\legend{DRF,SAF}
		\end{axis}
		\end{tikzpicture}
		\label{fig:gaussian3-memory}
	}
	
	\caption{Resource usages over time and execution time of SAF VS. DRF under different Gaussian workloads (\% per millisecond)}
	\label{fig:usage-gaussians}
\end{figure}

In the following, we compare the results of the SAF and DRF tests on different workloads. In previous experiments, the workload was generated as a uniform distribution. We also test three workloads in the form of a Gaussian distribution with varying means for CPU and RAM. Fig.~\ref{fig:usage-gaussians} shows the comparison of the SAF and DRF resource usage under the workload with Gaussian distributions with different CPU and RAM means.

\newgeometry{left=3cm,bottom=0.1cm}
\begin{figure} [H] 	%figure 11
	\centering
	\subfigure[\bf Gaussian Workload \normalfont ($\bar{CPU}$ = 2.75)]{
		\begin{tikzpicture}  % A-1 cpu
		\pgfplotsset{height=0.35\textwidth,width=0.51\textwidth,compat=newest,every tick label/.append style={font=\small}}
		\begin{axis}[
		xtick={50,150,...,1050}, ytick={1,2,...,7},xticklabel style={rotate=90},
		xlabel={Number of tasks},  ylabel=\bf{CPU usage over time},
		legend style={line width=0.2pt,font=\small}
		]
		\addplot [blue,mark=star] 
		coordinates {(50, 1.3) (150,1.9) (250,2.17) (350,2.4) (450,3.58) (650,2.4)  (850,2.43) (1050,2.38)};
		
		\addplot [purple,mark=star]
		coordinates {(50, 3.48) (150,3.7) (250,4.58) (350,5.72) (450,6.75) (650,4.3)  (850,4.13) (1050,3.26)};
		\legend{DRF,SAF}
		\end{axis}
		\end{tikzpicture}
		\label{fig:gaussian1-cpu}
	}
	%--------------------------------------------------------------
	\subfigure[\bf Gaussian Workload \normalfont ($\bar{RAM}$ = 575)]{
		\begin{tikzpicture} A-2 memory
		\pgfplotsset{height=0.35\textwidth,width=0.51\textwidth,compat=newest,every tick label/.append style={font=\small}}
		\begin{axis}[
		xtick={50,150,...,1050}, ytick={5,10,...,45},xticklabel style={rotate=90},
		xlabel={Number of tasks},  ylabel=\bf{Memory usage over time},
		legend style={line width=0.2pt,font=\small},ymin=0
		]
		
		\addplot [blue,mark=star] 
		coordinates {(50, 6.78) (150,10.51) (250,12.35) (350,14.06) (450,14.97) (650,13.58)  (850,13.73) (1050,13.43)};
		
		\addplot [purple,mark=star]
		coordinates {(50, 17.77) (150,20.31) (250,26.20) (350,33.17) (450,29.39) (650,23.94)  (850,23.21) (1050,18.26)};
		
		\legend{DRF,SAF}
		\end{axis}
		\end{tikzpicture}
		
		\label{fig:gaussian1-memory}
	}
\end{figure}
%********************************************************************************
\begin{figure} [H] 	%figure 11
	\centering
	\subfigure[\bf Gaussian Workload \normalfont ($\bar{CPU}$ = 4.5)]{
		\begin{tikzpicture}  % B-1 cpu
		\pgfplotsset{height=0.35\textwidth,width=0.51\textwidth,compat=newest,every tick label/.append style={font=\small}}
		\begin{axis}[
		xtick={50,150,...,1050}, ytick={1,2,...,7},xticklabel style={rotate=90},
		xlabel={Number of tasks},  ylabel=\bf{CPU usage over time},
		legend style={line width=0.2pt,font=\small}
		]
		
		\addplot [blue,mark=star] 
		coordinates {(50, 1.59) (150,2.37) (250,2.97) (350,3.46) (450,3.58) (650,3.43)  (850,3.42) (1050,3.3)};
		
		\addplot [purple,mark=star]
		coordinates {(50,4.39) (150,5) (250,6.61) (350,7.21) (450,6.75) (650,5)  (850,5.11) (1050,5.01)};
		\legend{DRF,SAF}
		\end{axis}
		\end{tikzpicture}
		\label{fig:gaussian2-cpu}
	}
	%--------------------------------------------------------------
	\subfigure[\bf Gaussian Workload \normalfont ($\bar{RAM}$ = 1050)]{
		\begin{tikzpicture} B-2 memory
		\pgfplotsset{height=0.35\textwidth,width=0.51\textwidth,compat=newest,every tick label/.append style={font=\small}}
		\begin{axis}[
		xtick={50,150,...,1050}, ytick={5,10,...,35},xticklabel style={rotate=90},
		xlabel={Number of tasks},  ylabel=\bf{Memory usage over time},
		legend style={line width=0.2pt,font=\small}
		]
		
		\addplot [blue,mark=star] 
		coordinates {(50, 5.67) (150,10.53) (250,12.66) (350,14.11) (450,14.97) (650,14.06)  (850,13.76) (1050,13.54)};
		
		\addplot [purple,mark=star]
		coordinates {(50,15.38) (150,22.07) (250,28.75) (350,29.54) (450,29.39) (650,21.01)  (850,20.68) (1050,20.61)};
		
		\legend{DRF,SAF}
		\end{axis}
		\end{tikzpicture}
		\label{fig:gaussian2-memory}
	}
	
\end{figure}
%********************************************************************************
\restoregeometry
%********************************************************************************
In all cases, the CPU's and RAM's standard deviation are 1.16 and 316.6, respectively. As it is clear, there is a considerable difference between the resource utilization of SAF and DRF in all Gaussian workloads. More precisely, in Fig.~\ref{fig:gaussian1-cpu}, the CPU utilization improvement is at its best with 50 tasks and the worst with 1050 tasks by 167\% and 37\% growth. In Fig.~\ref{fig:gaussian1-memory}, the RAM utilization improves in both mentioned cases by 161\% and 36\% in SAF compared to DRF. In Fig.~\ref{fig:gaussian2-cpu} and Fig.~\ref{fig:gaussian2-memory}, in the best state with 50 tasks, CPU and RAM utilization in SAF enhances by 175\% and 171\%, and at the worst case with 650 tasks, it increases by 45\% and 49\% respectively. Finally, in Fig.~\ref{fig:gaussian3-cpu} and Fig.~\ref{fig:gaussian3-memory}, the maximum, and minimum of increment occur in cases with 50 and 850 tasks. Thus in the case with 50 tasks, CPU and RAM utilization were improved by 130\% and 140\%. Also, in the case with 850 tasks, resource utilization improvement is 31\% and 30\%. To summarize, SAF improves CPU and RAM utilization on average by 92\% and execution time by 80\% compared to DRF.

\section{Conclusion}
\label{sec:8}
Considering the trade-off between fairness and performance in terms of resource utilization of the DRF scheduling policy, we decided to solve this issue using a long-term approach in task selection (entropy) and meta-heuristic algorithms like simulated annealing. In this paper, we presented a better scheduling policy, Simulated Annealing Fairness (SAF), for Hadoop YARN clusters that satisfies the goals of the DRF and improve resource utilization at the same time. As we mentioned, there is a trade-off between fairness and performance about resource utilization in the SAF, which our SAF scheduler can maintain a balance between them to reach better performance in Hadoop YARN scheduling. We implemented the SAF algorithm in Hadoop YARN v.3.1.0 and evaluated this scheme by running representative MapReduce benchmarks.

The results of our experiment demonstrated that the SAF improves performance in terms of both resource utilization and execution time by its long-term approach. Its superior performance relative to the DRF proves at least, 80\% increase in resource utilization (Memory and CPU) and 36\% reduction in MakeSpan and 70\% reduction in allocation time cost. It can also guarantee fairness for users since it uses concepts of fair policy.  

\section{Future Work}
\label{sec:9}

Hadoop YARN registers the Information regarding the job executions in the cluster as log files. This information could help the scheduler to decide faster when a new request arrives. Sometimes the resource requests from users are set to a much higher amount than the actual job's resource usage. It is better to check the upcoming jobs with their against these models and predict their actual resource requests.
For our future work, we intend to construct a recurrent neural network (RNN) model for jobs based on their information like resource request, resource usage, job type, and in case of an input file, the size of the input files to find a correlation between these features and execution time. Our goal is to predict the actual resource usage for upcoming jobs based on this model and prioritize them.
We intend to automate the cooling parameter value, which is the most crucial part of our scheduling scheme by designing a model based on the previous size and number of running jobs in the cluster to predict the cooling parameter in our scheduling policy future.

\bibliographystyle{spmpsci}      % mathematics and physical sciences
\bibliography{SAF}   % name your BibTeX data base

%\begin{thebibliography}{}
%%
%% and use \bibitem to create references. Consult the Instructions
%% for authors for reference list style.
%\end{thebibliography}

\end{document}